\documentclass[preprint,aps,amsmath,nofootinbib,tightenlines,superscriptaddress]{revtex4}
\usepackage{graphicx}
\usepackage{bm}
\usepackage{epsfig}

\newif\ifpdf
\ifx\pdfoutput\undefined
\pdffalse 
\else
\pdfoutput=1 
\pdftrue
\fi

\def\Bslash{B\!\!\!\!\slash}
\def\Dslash{D\!\!\!\!\slash}

\def\nslash{n\!\!\!\slash}
\def\bnslash{\bar n\!\!\!\slash}

\def\vslash{v\!\!\!\slash}
\def\OMIT#1{}

\newcommand{\nn}{\nonumber} 

\newcommand{\bn}{{\bar n}}
\newcommand{\bea}{\begin{eqnarray}}
\newcommand{\eea}{\end{eqnarray}}

\newcommand{\bnP}{\bar {\cal P}}

\newcommand{\mcdot}{\!\cdot\!}

\newcommand{\SCETa}{\mbox{${\rm SCET}_{\rm I}$ }}
\newcommand{\SCETb}{\mbox{${\rm SCET}_{\rm II}$ }}

\newcommand{\lvec}[1]{\kern0.166666em\vbox{\hbox{\kern-0.05em\lower 0.4 em\hbox{\scriptsize$\leftarrow$}}\hbox{$ #1 $}}\hbox to -0.0em{}}

\begin{document}
\ifpdf
\DeclareGraphicsExtensions{.pdf, .jpg}
\else
\DeclareGraphicsExtensions{.eps, .jpg}
\fi


\preprint{ \vbox{ \hbox{hep-ph/0405290} \hbox{MIT-CTP-3497} }}

\title{\phantom{x}
\vspace{0.5cm}
Heavy Quark Symmetry in Nonleptonic\\[3pt]
B-Decays to Excited Charmed Mesons
\vspace{0.6cm}
}

\author{Sonny Mantry}
\affiliation{Center for Theoretical Physics, Massachusetts Institute for
Technology,\\ Cambridge, MA 02139\footnote{Electronic address: mantry@mit.edu}
\vspace{0.2cm}}

\vspace{0.2cm}
\vspace{0.3cm}

\begin{abstract}
\vspace{0.3cm}

We show in a model independent way the equality of the branching fractions and strong
phases for $\bar B\to D_1M$ and 
$\bar B\to D_2^*M$ at leading order in $\Lambda _{QCD}/Q$ and $\alpha _s(Q)$ where $Q=\{m_b,m_c,E_M\}$
and $M$ is a light meson. These results apply in the color allowed and color suppressed
channels and
follow from a factorization theorem in SCET combined with
heavy quark symmetry. The expected heavy quark symmetry suppression of leading 
order contributions
in the color allowed sector
based on analysis of semileptonic decays, 
is shown to disappear at maximum recoil. Subleading corrections are
suppressed by at least one power of $\Lambda _{QCD}/Q$ and this is explicitly
verified for subleading semileptonic form factors at maximum recoil.

\end{abstract}

\maketitle

\section{Introduction} \label{sect_intro}

Two-body non-leptonic $B$-decays have generated considerable theoretical
interest~\cite{BSW,DG,phen,BlSh,BuSi,PW,bbns,Ligeti, bps, xing0,NePe,Ben,Rosner,
xing,Li,Bauer:2004tj} in recent years. Most recently, Soft Collinear Effective 
Theory(SCET)~\cite{SCET,bps2} has been used with remarkable success in 
understanding such decays. Using SCET, $B$-decays of the type 
$\bar B\to D^{(*)}M$~\cite{Mantry} where $M=\pi ,\rho ,K , K^*$ have been
studied in quite some detail. In this paper, we
investigate such decays when the final state charmed meson is in an orbitally
excited state such as the $D_1$ and $D_2^*$(see Table ~\ref{table_hqs})
collectively referred to as $D^{**}$. 

$\bar B\to D^{**}K$ decays have been recently proposed~\cite{Sinha:2004ct} as 
candidates for a theoretically clean extraction of the CKM angle $\gamma $
making such decays all the more interesting to study. These decays also raise
interesting questions regarding the power counting scheme used to make quantitative 
phenomenological predictions.
Based on analysis of semileptonic decays~\cite{Leibovich:1997em} near zero recoil,
the leading order contributions are expected to be suppressed due to heavy
quark symmetry constraints. This suggests that subleading contributions could
have a significant effect on leading order predictions in $\bar B\to D^{**}M$ type
processes.
We will address these issues on power counting and provide a resolution.
On another note, the $\bar B^0\to D^{(*)0}\rho ^0$ rates are more
difficult
to extract cleanly from experimental data due to background contributions from intermediate
$D^{**}$ states. In particular, in the $D^{*0}$ channel only an upper bound on the
branching fraction has been measured~\cite{PDG} and the errors in the $D^0$ channel are
still fairly large~\cite{Belle}. This has made it difficult
to test the SCET prediction~\cite{Mantry} relating the $D$ and $D^*$ amplitudes.
With the $\rho ^0$ meson primarily decaying to $\pi
^+\pi ^-$ and the excited $D^{**+}$ mesons decaying to $D^{(*)0}\pi ^+$
the same final state is observed for $\bar B^0\to D^{**+}\pi ^-$
and $\bar B^0\to D^{(*)0}\rho ^0$. Thus, a precise extraction of the 
$\bar B^0\to D^{(*)0}\rho ^0$ rates requires us to better understand 
$B$ decays to excited charmed mesons. 

\begin{figure}[!t]
\vskip0.1cm
 \centerline{
  \mbox{\epsfxsize=4.5truecm \hbox{\epsfbox{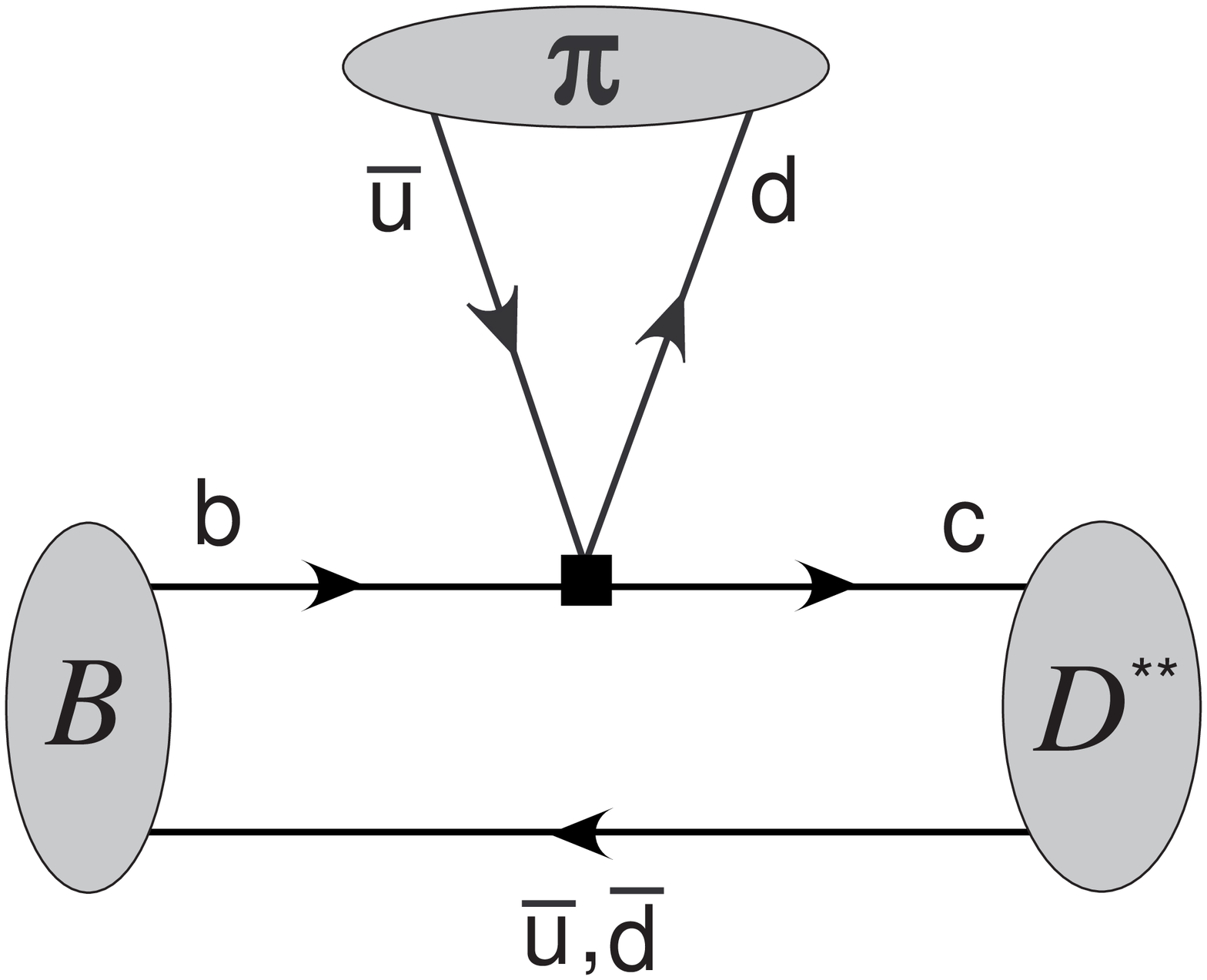}} }
   \hspace{0.7cm}
  \mbox{\epsfxsize=4.5truecm \hbox{\epsfbox{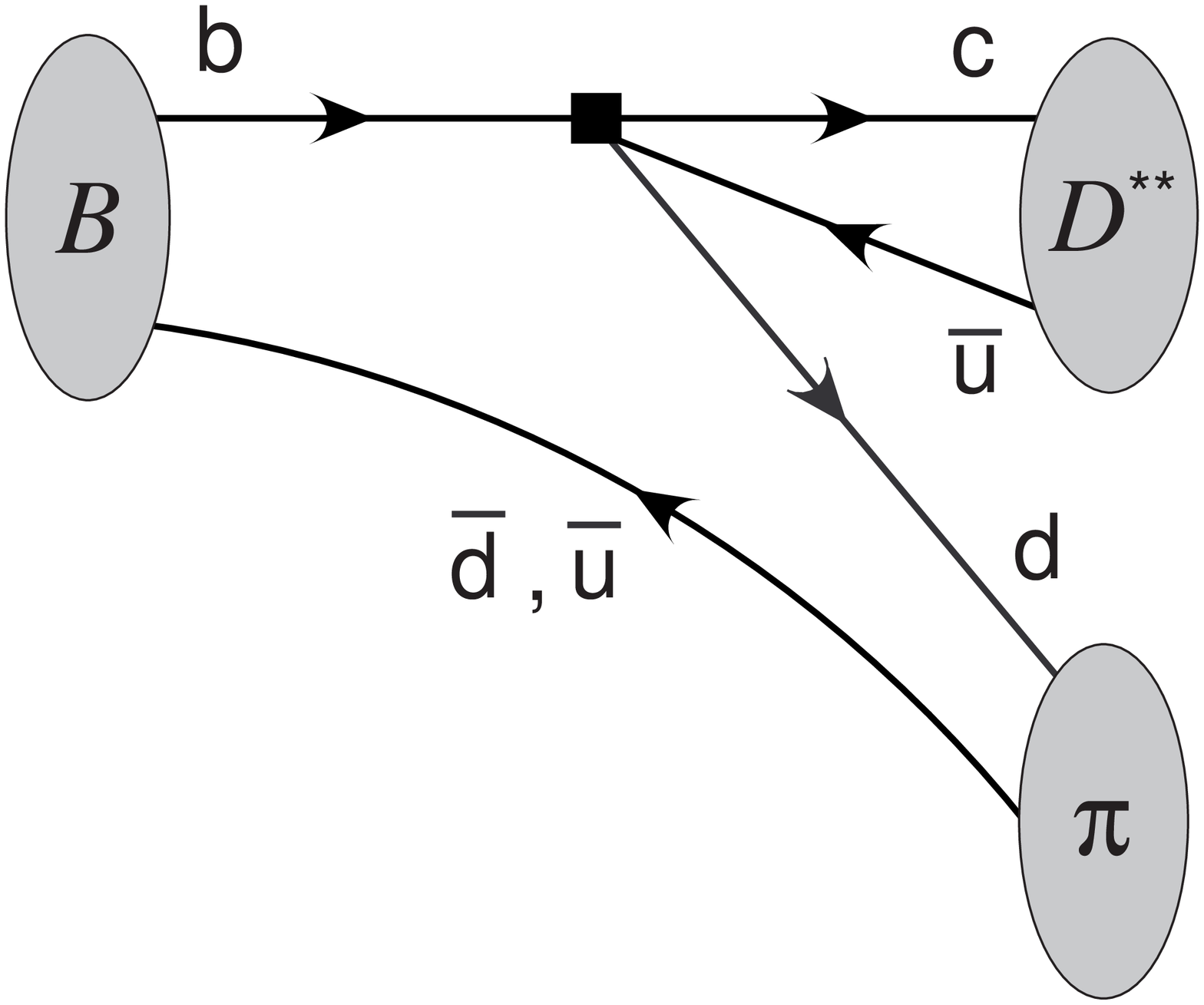}} }
   \hspace{0.7cm}
  \mbox{\epsfxsize=4.5truecm \hbox{\epsfbox{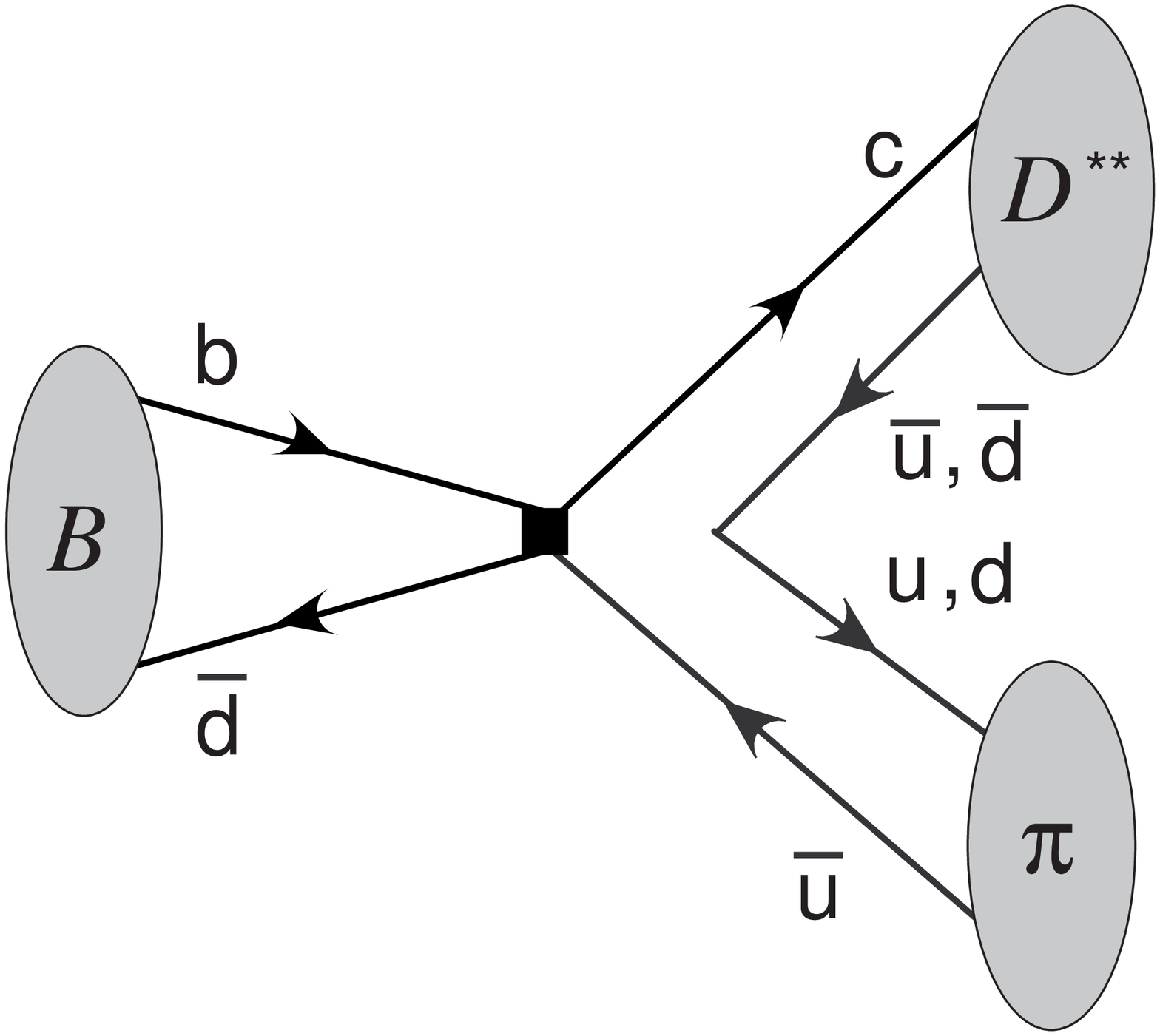}} }
  } 
 \vspace{0.2cm}
 \centerline{ 
  \vbox{\raisebox{-0.15cm}{T:}\hspace{0.3cm}
        $\bar B^0\to D^{**+}\pi^-$ \hspace{2.5cm} 
        \raisebox{-0.15cm}{C:}\hspace{0.3cm}
        $ B^-\to D^{**0}\pi^-$ \hspace{2.1cm}
        \raisebox{-0.15cm}{E:}\hspace{0.3cm}
        $\bar B^0\to D^{**+}\pi^-$  \nn \\ 
        \hspace{0.55cm}
        $B^-\to D^{**0}\pi^-$ \hspace{3.4cm}
        $\bar B^0\, \to D^{**0}\pi^0$ \hspace{3.05cm}
        $\bar B^0 \to D^{**0}\pi^0$ \hspace{0.1cm}}
 } 
\vskip-0.3cm
\caption[1]{Topologies contributing to $\bar B\to D^{**}M$ decays
where $M$ is a light meson. The diagrams are shown for the 
simplest case of the $D^{**}\pi$ final state.}
\label{fig_qcd} 
\vskip0cm
\end{figure}

The $\bar B\to (D^{(*)},D^{**})M$ decays proceed via
three possible topologies shown in Fig.~\ref{fig_qcd} for the case of a $D^{**}\pi$
final state. In the $T$(tree) topology
the light meson is produced directly at the weak vertex. In the $C$(color
suppressed) and $E$(W-exchange or weak annihilation) topologies, a spectator quark
ends up in the final state light meson. The $C$ and $E$ topologies are
suppressed in the large $N_c$ limit. The $\bar B^0\to D^{(*)0}M^0$ type decays 
that proceed exclusively through these $C$ and $E$ topologies will be generically 
referred to as the color suppressed modes. Other decay channels that are 
dominated by the $T$ topology will be referred to as the color allowed modes.

We review some of the recent theoretical and experimental results for the 
above mentioned decays. 
Factorization theorems~\cite{Mantry,bps} in SCET
have been proven for $\bar B\to D^{(*)}M$ decays
(including the often less understood
color suppressed modes)to first non-vanishing order in 
$\Lambda _{QCD}/Q$. Here $Q$ is a hard scale on the order of the bottom and 
charmed quark masses $m_{b,c}$ or the light meson energy $E_M$. It was
shown that the $C$ and $E$ topologies are suppressed by $\Lambda _{QCD}/Q$
relative to $T$ which explains the observed~\cite{PDG} suppression of the 
$\bar B^0\to D^{(*)0}M^0$
color suppressed modes. Using heavy quark 
symmetry in conjunction with factorization, quantitative 
model independent phenomenological results relating the $D$ and $D^*$ amplitudes
were obtained 
\begin{eqnarray}\label{RM}
  \frac{Br(\bar B \to D^{*}M)}
    {Br(\bar B \to DM)} = 1 \,.
\end{eqnarray} 
These results are to leading order in $\alpha _s(Q)$ and $\Lambda _{QCD}/Q$ and 
hold true in the color allowed channels for all the above mentioned light mesons
$M$~\cite{PW} . For 
the color suppressed modes the predictions hold for 
$M=\pi , \rho ,K, K^*_{||}$~\cite{Mantry}. For the kaons in the color suppressed channels, there 
are additional 
non-perturbative functions from long distance operators which 
require the
$K^*$s to be longitudinally polarized for the prediction to hold. Predictions of the type in Eq.~(\ref{RM}) have also
been made for the case of the baryon decays~\cite{Leibovich:2003tw} 
$\Lambda _b\to \Xi _c^{(*)}M$ and $\Lambda _b\to \Sigma _c^{(*)}M$ where heavy
quark symmetry relates $\Xi _c$ to $\Xi _c^*$ and 
$\Sigma _c$ to $\Sigma _c^*$. In this case the ratio of branching fractions 
was found to be $2$.
\begin{table}[t!]
\begin{center}
\begin{tabular}{|c|c|c|c|c|c|c|}
\hline
Mesons & $s_l^{\pi _l}$ & $J^P$ & $\bar m$(GeV) \\
\hline\hline
$(D,D^*)$ & $\frac {1}{2}^-$ & $(0^-,1^-)$ & $1.971$ \\
$(D_0^*,D_1^*)$ & $\frac {1}{2}^+$ & $(0^+,1^+)$ & $2.40$ \\
$(D_1,D_2^*)$ & $\frac {3}{2}^+$ & $(1^+,2^+)$ & $2.445$ \\
\hline\hline
\end{tabular}
\end{center}
\caption{The HQS
doublets are labeled by $s_l^{\pi _l}$. Here $s_l$ denotes the spin of
the light degrees of freedom and $\pi _l$ the parity. The $D,D^*$ mesons are
$L=0$ negative parity mesons. The $D_0^*,D_1^*$ and $D_1,D_2^*$ are excited
mesons with $L=1$ and positive parity. $\bar m$ refers to the average mass of
the HQS doublet weighted by the number of helicity
states~\cite{Leibovich:1997em}.\label{table_hqs}} 
\end{table}

The relation between the $D$ and $D^*$ amplitudes in Eq.~(\ref{RM}) can be understood in 
terms of soft-collinear factorization and Heavy Quark
Symmetry(HQS)~\cite{Isgur:vq,Isgur:ed}. 
In the heavy quark limit $m_c \to \infty $ and in the absence of hard gluons, 
the $D$ and $D^*$ charmed mesons sit in a HQS doublet $(D,D^*)$.  Members within a HQS doublet are
distinguished by the coupling of the spin of the charmed quark
with the spin of the light degrees of freedom ($s_l$)~\cite{hbook}. Their total spin 
is given by $J_{\pm } = s_l \pm \frac {1}{2}$. Since 
spin dependent chromomagnetic interactions are $\Lambda _{QCD}/m_c$  
suppressed, the $D$ and $D^*$ states are degenerate at leading order. 
However, the presence of collinear gluons in the energetic light meson can spoil
HQS since chromomagnetic corrections of order $E_M/m_c$ from such gluons are not
suppressed. Factorization of these collinear modes from the soft degrees of
freedom becomes crucial in restoring this symmetry. SCET provides us with such a
factorization theorem where the amplitude factors into soft 
$\langle D^{(*)}|\cdots |B\rangle $ and collinear $\langle M|\cdots |0\rangle $
matrix elements at leading order. One is now free to apply HQS in the soft sector and
the result of Eq.~(\ref{RM}) is a statement of this symmetry. 

There exists a tower of HQS doublets for the charmed mesons where $(D,D^*)$ sits
at the base. The first three HQS doublets are listed in
Table~\ref{table_hqs}. In this paper we extend the analysis to the case where the 
final state
charmed mesons are $D_1$ or $D_2^*$ which sit in the third HQS doublet. 
A similar analysis can be done for the $(D_0^*,D_1^*)$ doublet but
these are difficult to observe due to their relatively broad width~\cite{hbook}. For
this reason, we restrict our analysis to the $(D_1,D_2^*)$ doublet. 
\begin{figure}[!t]
\vskip-0.1cm
\includegraphics[height=.23\textheight]{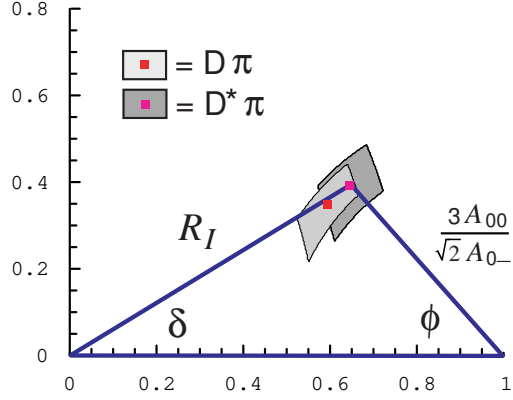}
\vskip-0.3cm
\caption[1]{
  The ratio of isospin amplitudes $R_I = A_{1/2}/(\sqrt2 A_{3/2})$ and strong
  phases $\delta$ and $\phi$ in $\bar B\to D\pi$ and $\bar B\to D^*\pi$ taken
  from Ref.~\cite{Mantry:2004qg}. The
  central values following from the $D$ and $D^*$ data in Table I are denoted by
  squares, and the shaded regions are the $1\sigma$ ranges computed from the
  branching ratios. The overlap of the $D$ and $D^*$ regions shows that the 
  prediction in Eq.~(\ref{RM}) works well. }
\label{fig_isospin} 
\vskip0cm
\end{figure}

For $M=\pi , \rho $ the final state can be decomposed into an isospin 
$I=1/2,3/2$ basis allowing us to parametrize the physical amplitudes in terms 
of the isospin amplitudes $A_{1/2}$ and $A_{3/2}$

\begin{eqnarray} \label{As}
  A_{+-} = A(\bar B^0\, \to D^+ \pi^-) &=& \frac{1}{\sqrt3} A_{3/2} + 
  \sqrt{\frac23} A_{1/2} = T+E   \,,\nn   \\
  A_{\,0-} = A( B^- \to D^0\, \pi^-) &=& \sqrt3 A_{3/2} = T+C  \,,\nn  \\
  A_{\,00\,} = A(\bar B^0\, \to D^0\, \pi^0\,) &=& \sqrt{\frac23} A_{3/2} -
  \frac{1}{\sqrt3} A_{1/2} = \frac{1}{\sqrt2}(C-E)\,.
\end{eqnarray}
These relations also hold true for $D^{**}$ mesons in the final state. 
The relative phase between the isospin amplitudes is defined as 
$\delta={\rm arg}(A_{1/2} A_{3/2}^*)$. The above equations can be brought into
the form:
\begin{eqnarray}\label{triangle}
R_I+\frac {3A_{00}}{\sqrt {2}A_{0-}}=1,
\end{eqnarray}
where $R_I = A_{1/2}/(2 A_{3/2})$. This relation can be expressed as a triangle
in the complex plane as shown in Fig.~\ref{fig_isospin} for the $\bar B\to
D^{(*)}\pi $ channels. The overall phase was chosen so that 
the $A_{0-}$ amplitude is real. The angle 
$\phi $ is the non-perturbative 
strong phase of the color suppressed amplitude $A_{00}$. A novel mechanism for the
generation of this phase was discussed in ~\cite{Mantry}. The prediction 
of Eq.~(\ref{RM}) manifests itself as the identical overlap of the $D\pi$ and 
$D^*\pi$ isospin
triangles i.e. $\delta =\delta ^*$ and $\phi =\phi ^*$. Fig.~\ref{fig_isospin} shows 
the remarkable agreement of this prediction with data. For the $\bar B\to
D^{**}M$  modes not enough data exists to construct analogous isospin
triangles. However, the experimental scene is quite active with most recent
measurements in the color allowed sector giving the ratio
\begin{eqnarray}\label{Belle}
\frac {Br(B^- \to D_2^{*0}\pi ^-)}{Br(B^- \to D_1^0\pi ^-)} = 0.79 \pm 0.11,
\end{eqnarray}
obtained after averaging the Belle~\cite{Abe:2003zm} and Babar~\cite{Aubert:2003hm} data.
In this paper, we shed light on this ratio and also make predictions
in the color suppressed sector.

In extending the analysis to include excited charmed mesons, the 
constraint of HQS 
introduces possible complications in the power counting scheme. 
HQS requires the matrix elements of the weak current between $B$ 
and $(D_1,D_2^*)$ 
to vanish at zero recoil~\cite{Isgur:wq}. This requires that they be proportional to 
some positive power of $(\omega -1)$ at leading order in $\Lambda _{QCD}/Q$.
Here $\omega =v\cdot v'$ where $v$ and $v'$ are the velocities of the 
bottom and charm quarks respectively and $v^2=v^{'2}=1$. 
For semileptonic decays this means that HQS breaking 
$\Lambda _{QCD}/Q$ corrections can compete with the leading order
prediction~\cite{Leibovich:1997em,Leibovich:2003tw}. For example, if the amplitude were to have the generic form
\begin{eqnarray}\label{HQSconstraint}
A(\omega )\sim (\omega -1)[1 + \Lambda _{QCD}/Q +\cdots ] + [0 + \Lambda _{QCD}/Q
+\cdots ], 
\end{eqnarray}
and $(\omega -1)\sim \Lambda _{QCD}/{Q}$, then the subleading $\Lambda _{QCD}/{Q}$
terms in the second square bracket 
are of the same order as the leading order terms in the first square 
bracket. The effect of the subleading
corrections is especially important near zero recoil where $\omega
\to 1$. The two body decays $\bar B\to (D_1,D_2^*)M$ occur at maximum recoil 
where 
$(\omega _0-1)\sim 0.3$ which is numerically of the same order as 
$\Lambda _{QCD}/{Q}$. One is thus forced to consider the role of
subleading corrections and how they compare with the leading order
predictions. However, we will see that maximum recoil is a
special kinematic point at which the constraint of HQS enters in a 
very specific manner so as to
preserve the $\Lambda _{QCD}/Q$ power counting scheme. The main results of this paper
are
\begin{itemize}

\item At leading order, the ideas of factorization, generation of 
non-perturbative 
strong phases, and the relative 
$\Lambda _{QCD}/Q$ suppression of the color suppressed modes are the same
for $B$-decays to excited charmed mesons $\bar B\to D^{**}M$ and to 
ground state charmed mesons $\bar B\to D^{(*)}M$.

\item The constraint of HQS takes on a 
different character at maximum
recoil compared to expectations from the analysis of semileptonic 
decays near zero recoil. In particular, at maximum recoil 
there is no suppression of the leading order contribution due to HQS.
Thus, the SCET/HQET power counting scheme remains intact and allows us to rely 
on leading order predictions up to corrections suppressed by 
at least $\Lambda _{QCD}/Q$. We verify this explicitly for subleading
corrections to the semileptonic form factors at maximum recoil.

\item At leading order, factorization combined with HQS predicts the equality 
of the $\bar B\to D_1M$ and $\bar B\to D_2^*M$ branching fractions and their
strong phases. In the color suppressed sector, this prediction is quite non-trivial 
from the point of naive factorization since the tensor meson $D_2^*$ cannot be created
via a V-A current.

\item Recent data ~\cite{Abe:2003zm, Aubert:2003hm} reports a 20\% deviation of the ratio of
branching fractions
from unity in the color allowed sector. The subleading corrections of order $\Lambda_{QCD}/Q$ are 
expected to be of this same size and could explain this deviation from unity. 
\end{itemize}

In section~\ref{sect_data} we review the analysis of SCET
for $\bar B\to D^{(*)}M$. At leading order, most of 
the results can be identically carried over to the analysis for the 
excited
HQS doublet $(D_1,D_2^*)$. In section~\ref{sect_ECM} we consider
the effect of subleading corrections and phenomenological
results are given in section IV.







\section{Soft Collinear Effective Theory for $\bar B\to D^{(*)}M$} \label{sect_data}

Observing the decay in the rest frame of the $B$ meson, one can identify two 
types of degrees of freedom with offshellness $p^2\sim \Lambda _{QCD}^2$
that are responsible for binding the hadrons. These are the
collinear $(p^+,p^-,p_\perp )\sim Q(\eta ^2,1,\eta )$ and 
soft $(p^+,p^-,p_\perp )
\sim Q(\eta ,\eta ,\eta )$ degrees of freedom where 
$\eta \sim \Lambda _{QCD}/Q$.
The formalism of SCET allows us to construct an effective theory of this process
directly in terms of these relevant soft and collinear modes with all other
offshell modes integrated out. This effective theory at the hadronic scale 
is given the name \SCETb \footnote{The soft-collinear messenger modes of 
Ref.~\cite{Becher:2003qh} could play a role in subleading corrections which we
will not consider. The nature of these messenger modes is still unclear due to their dependence
on the choice of infrared regulator~\cite{Beneke:2003pa,Bauer:2003td}.}.

The $\bar B\to D^{(*)}M$ processes receive contributions from 
various effects 
occuring at different distance scales. A complete description of 
these decays requires us to flow between effective theories from the electroweak 
scale down to the hadronic scale. Each effective theory along the way 
contributes the neccessary mechanism for the decay to proceed. These mechanisms
are encoded as effective operators with appropriate Wilson coefficients 
in the next effective theory on our way down to \SCETb at the hadronic scale. 

The $b\to c$ quark flavor changing process occurs at 
the electroweak scale $(p^2 \sim m_W^2)$ through a $W$-exchange process. 
The $W$ boson is then integrated out to
give the effective Hamiltonian
\begin{eqnarray}\label{Hw}
 {\cal H}_W = \frac{G_F}{\sqrt2} V_{cb} V_{ud}^* [ C_1(\mu)
 (\bar c b)_{V-A} (\bar d u)_{V-A} + C_2(\mu)
 (\bar c_i b_j)_{V-A} (\bar d_j u_i)_{V-A} ]\,,
\end{eqnarray}
where $i,j$ are color indices, and for $\mu_b=5\,{\rm GeV}$, $C_1(\mu_b)= 1.072$
and $C_2(\mu_b)=-0.169$ at NLL order in the NDR scheme~\cite{Buras}. 
This Hamiltonian gives rise to the three distinct topologies
through which the decay can proceed as shown
in Fig.~\ref{fig_qcd}.

Next we would like to match $H_W$ onto operators in \SCETb with soft and
collinear degrees of freedom. However, the
soft-collinear interactions produce offshell modes $p^2\sim Q\Lambda _{QCD}$
that are not present in \SCETb. These modes have momentum scalings 
$(p^+,p^-,p_\perp )\sim Q(\eta ,1,\eta )$ and have to be integrated
out~\cite{bps2}.
Instead, it becomes more convenient to go through an intermediate effective 
theory \SCETa ~\cite{bps4}
at the scale $Q\Lambda _{QCD}$ and do the matching in two steps.
\SCETa is a theory of ultrasoft 
$(p^+,p^-,p_\perp)\sim Q(\lambda ^2,\lambda ^2,\lambda ^2)$
and hard-collinear $(p^+,p^-,p^\perp)\sim Q(\lambda ^2,1,\lambda )$ modes 
where $\lambda = \sqrt \eta = \sqrt\frac {\Lambda _{QCD}}{Q}$. The ultrasoft  
modes are identical to the soft modes and the hard-collinear modes play 
the role of the offshell modes produced by the soft-collinear
interactions in \SCETb. The hard-collinear modes are eventually matched 
onto the collinear modes of \SCETb.  
This two step
matching procedure allows us to avoid dealing directly with non-local 
interactions, altough it is also possible to construct \SCETb directly from
QCD~\cite{HN}.
In summary, one arrives at the effective theory \SCETb at the hadronic scale 
through a series of matching and running procedures starting with the 
Standard Model(SM) 

\begin{eqnarray}\label{match}
SM \to H_W \to \SCETa \to \SCETb .\nonumber
\end{eqnarray}
In the above chain of effective theories, each matching calculation introduces
Wilson coefficients which encode the physics of harder scales. These ideas are 
summarized in Table ~\ref{matchtable} and are illustrated in Fig. 3. We now
briefly review the details of the procedure just discussed. 
\begin{table}[t!]
\begin{center}
\begin{tabular}{|c|c|c|c|}

\hline
Theory & Scale & Wilson Coefficients & Physics Effect \\
\hline\hline
SM & $\mu ^2 \sim m_W^2$ & - & $b\to c$ quark flavor transition \\
$H_W$ & $\mu ^2 \sim Q^2$ & $C_1,C_2$ & $W$ boson integrated out \\
\SCETa & $\mu ^2 \sim Q\Lambda _{QCD}$ & $C_L,C_R$ &soft-collinear transitions  \\
\SCETb & $\mu ^2 \sim \Lambda _{QCD}^2$ & $J$ & binding of hadrons \\
\hline\hline
\end{tabular}
\end{center}
{\caption{The effective theories at different distance scales and the effects
they provide for the $B\to DM$ process to occur. The Wilson coefficients that
show up in each theory are also given.}
\label{matchtable} }
\end{table}

\begin{figure}[h]
  \epsfxsize = 10cm
  \centerline{\epsfbox{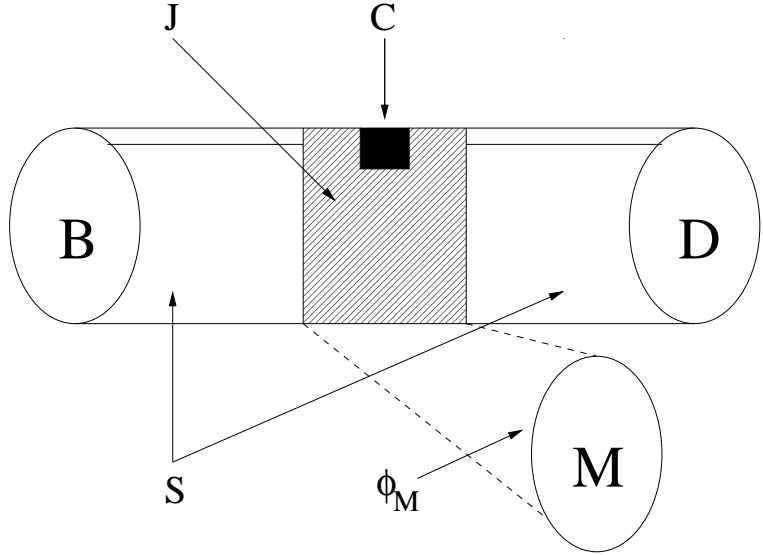}}
  \caption{A schematic representation of the $B \to DM$ process and the 
  contributions it receives from effects at different distance scales.
  The shaded black box is the weak vertex where the $b\to c$ transition takes
  place, the shaded grey region is where soft spectator quarks are converted to
  collinear quarks that end up in the light meson, and the unshaded regions are
  where non-perturbative processes responsible for binding of hadrons take
  place. These regions correspond to the functions $C$, $J$, $S$, and $\phi _M$
  as labeled in the figure. For the color allowed modes, where the 
  light meson is produced directly at the weak vertex and no soft-collinear
  transitions involving the spectator quarks are required, the jet function $J$
  is trivially just one. }
  \label{Scales}
\end{figure}

\subsubsection{Color Allowed Modes}

The leading order contribution to color allowed modes comes via the $T$ topology
(see Fig.~\ref{fig_qcd})where the final state light meson is emitted directly at 
the weak vertex.
After the $W$ boson is integrated out we arrive at the effective Hamiltonian
$H_W$ in Eq.~(\ref{Hw}). Next, we match $H_W$ onto \SCETa :
\begin{eqnarray} \label{QVI0}
  \sum_{1,2} C_i O_i \to 4 \sum_{j=L,R} \int\!\! d\tau_1 d\tau_2 \big[
  C^{(0)}_{j} (\tau_1,\tau_2)
  {\cal Q}^{(0)}_{j}(\tau_1,\tau_2) + C^{(8)}_{j} (\tau_1,\tau_2)
  {\cal Q}^{(8)}_{j}(\tau_1,\tau_2) \big]\,,
\end{eqnarray}
where the $O_i$ refer to the four quark operators in Eq.~(\ref{Hw}) and the
prefactor $\frac {G_F}{\sqrt 2}V_{cb}V^*_{ud}$ has been dropped from both sides
of the equation. 
At leading order in \SCETa there are four operators [$j=L,R$]
\begin{eqnarray}\label{QVI}
 {\cal Q}^{(0)}_{j}(\tau_1,\tau_2) &=& 
\big[\bar h_{v'}^{(c)} \Gamma^h_j  h_v^{(b)} \big] 
 \big[(\bar\xi_n^{(d)} W)_{\tau_1} \Gamma_n 
  (W^\dagger \xi_n^{(u)})_{\tau_2} \big] \,,\\
 {\cal Q}^{(8)}_{j}(\tau_1,\tau_2) &=&
\big[\bar h_{v'}^{(c)} Y \Gamma^h_j T^a Y^\dagger h_v^{(b)} \big] 
 \big[(\bar\xi_n^{(d)} W)_{\tau_1} \Gamma_n T^a 
 (W^\dagger \xi_n^{(u)})_{\tau_2} \big]\nn \,.
\end{eqnarray}
The superscript $(0,8)$ denotes the $1\otimes 1$ and $T^a\otimes T^a$ color
structures. The Dirac structures on the heavy side are \hbox{$\Gamma^h_{L,R}=
  \nslash P_{L,R}$} with $P_{R,L} = \frac12(1\pm \gamma_5)$, while on the
hard-collinear side we have $\Gamma_n = \bnslash P_L/2$. The momenta 
labels are
defined by $(W^\dagger \xi_n)_{\omega_2} = [\delta(\omega_2\!-\!\bnP)\:
W^\dagger \xi_n]$. The tree level matching conditions are:
\begin{eqnarray}
  C_L^{(0)}(\tau_i) &=& C_1 + \frac{C_2}{N_c} \,,\qquad
  C_L^{(8)}(\tau_i) = 2 C_2 \,,\qquad C_R^{(0,8)}(\tau_i) = 0\,.
\end{eqnarray}
Matching corrections of order ${\cal O}(\alpha_s)$ can be found in 
Ref.~\cite{bbns}.

The operators in Eq.~(\ref{QVI}) are written in terms of 
hard-collinear fields which
do not couple to usoft particles at leading order. This was achieved by a
decoupling field redefinition~\cite{bps2} on the hard-collinear fields $\xi_n\to
Y\xi_n$ etc.  The operators in Eq.~(\ref{QVI}) are then matched onto \SCETb to
give [$\omega_i=\tau_i$]
\begin{eqnarray}\label{QV}
 {\cal Q}^{(0)}_{j}(\omega_1,\omega_2) &=& 
\big[\bar h_{v'}^{(c)} \Gamma^h_j  h_v^{(b)} \big] 
 \big[(\bar\xi_n^{(d)} W)_{\omega_1} \Gamma_n 
  (W^\dagger \xi_n^{(u)})_{\omega_2} \big] \,,\\
 {\cal Q}^{(8)}_{j}(\omega_1,\omega_2) &=&
\big[\bar h_{v'}^{(c)} S \Gamma^h_j T^a S^\dagger h_v^{(b)} \big] 
 \big[(\bar\xi_n^{(d)} W)_{\omega_1} \Gamma_n T^a 
 (W^\dagger \xi_n^{(u)})_{\omega_2} \big]\nn \,,
\end{eqnarray}
where $W$ and $S$ are the hard-collinear and soft Wilson lines respectively.
As mentioned earlier, the 
hard-collinear fields of \SCETa in Eq.~(\ref{QVI}) match onto the 
collinear fields of \SCETb in Eq.~(\ref{QV}). The ultrasoft modes of \SCETa match 
onto the soft modes of \SCETb since these are identical. Note that since
the collinear quarks in the pion are produced directly at the weak vertex,
the jet functions $J^{(0,8)}$ which are the Wilson coefficients that arise in
matching \SCETa onto \SCETb are just one. 

At leading order in 
$\Lambda _{QCD}/Q$ only the operators ${\cal Q}^{(0)}_{L, R}$ and the leading 
order collinear and soft
Lagrangians (${\cal L}_{c}^{(0)}$, ${\cal L}_s^{(0)}$), contribute to the
$B^-\to D^{(*)0}\pi^-$ and $\bar B^0\to D^{(*)+}\pi^-$ matrix elements.  The
matrix elements of ${\cal Q}_{L,R}^{(8)}$ vanish because they factorize into a
product of bilinear matrix elements and the octet currents give vanishing
contribution between color singlet states~\cite{bps}. Since the soft and
collinear modes are decoupled at leading order, the matrix
elements of ${\cal Q}^{(0)}_{L, R}$ factorize into a soft $B\to D$ matrix
element and a light cone pion wave function. By employing the trace formalism
of HQET on the soft sector, the factorized amplitude
for the color allowed modes to all orders in $\alpha _s(Q)$ and to leading 
order in $\Lambda _{QCD}/Q$ can now be written as:

\begin{eqnarray} \label{factLO}
  A(B\to D^{(*)}\pi) = N^{(*)}\: \xi(w_0,\mu) \int_0^1\!\!dx\: 
   T^{(*)}(x,m_c/m_b,\mu)\: \phi_\pi(x,\mu)  \,,
\end{eqnarray}
where normalization factor is given by
\begin{eqnarray} \label{N}
  N^{(*)}= \frac{G_F V_{cb}^{\phantom{*}}V_{ud}^*}{\sqrt{2}}\:
   { E_\pi f_\pi} \sqrt{m_{D^{(*)}} m_B} \: 
    \Big( 1+\frac{m_B}{m_{D^{(*)}}} \Big) \,.
\end{eqnarray}
$\phi_\pi(x,\mu)$ is the non-perturbative pion light cone wave 
function, and $\xi(w_0,\mu)$ is the Isgur-Wise function evaluated at maximum
recoil. The hard
coefficient $T^{(*)}(x,\mu)=C^{(0)}_{L\pm R}((4x-2)E_\pi,\mu,m_b)$, where the
$\pm$ correspond to the $D$ and $D^*$ respectively, and $C^{(0)}_{L\pm R}=
C^{(0)}_{L} \pm C^{(0)}_{R}$.
With $C_R^{(0)}=0$ at leading order in $\alpha _s(Q)$ we have
$T(x,\mu)=T^{*}(x,\mu)$. In addition, in the heavy quark limit $N=N^*$ and one is
led to the result in Eq.~(\ref{RM}) for the color allowed modes. 

\subsubsection{Color Suppressed Modes}

\begin{figure}[!t]
\vskip-0.3cm
 \centerline{\hspace{0.6cm}
  \mbox{\epsfxsize=5.5truecm \hbox{\epsfbox{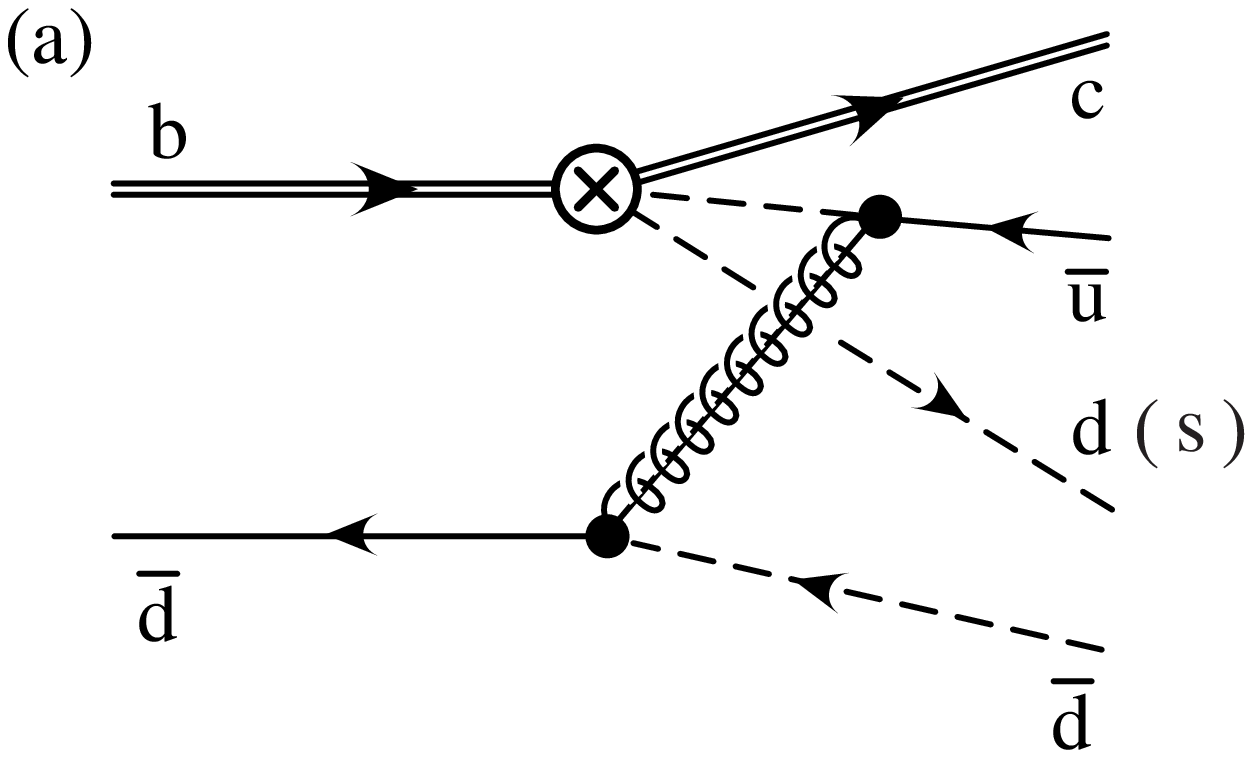}} }
  \hspace{0.cm}
  \raisebox{0.4cm}{\mbox{\epsfxsize=5.7truecm
      \hbox{\epsfbox{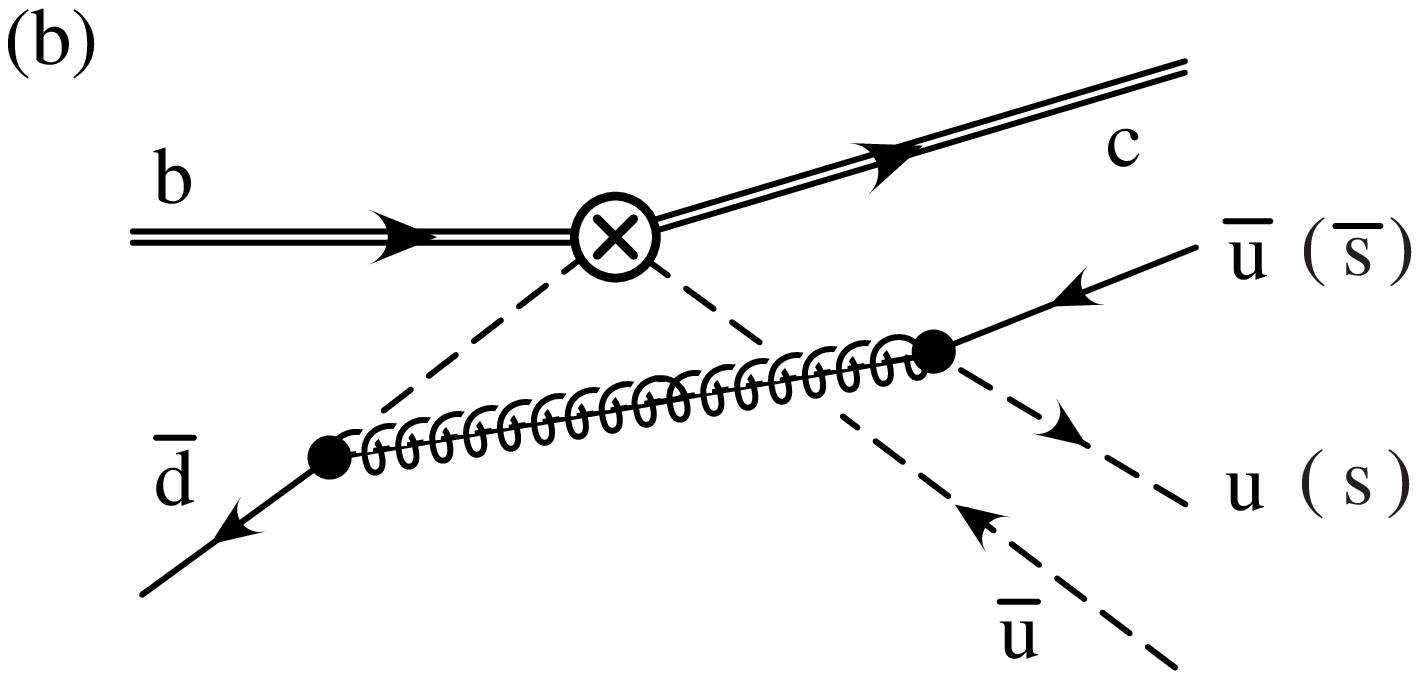}}} }
  }
\vskip-0.2cm
  \centerline{\hspace{0.6cm}
  \mbox{\epsfxsize=5.5truecm \hbox{\epsfbox{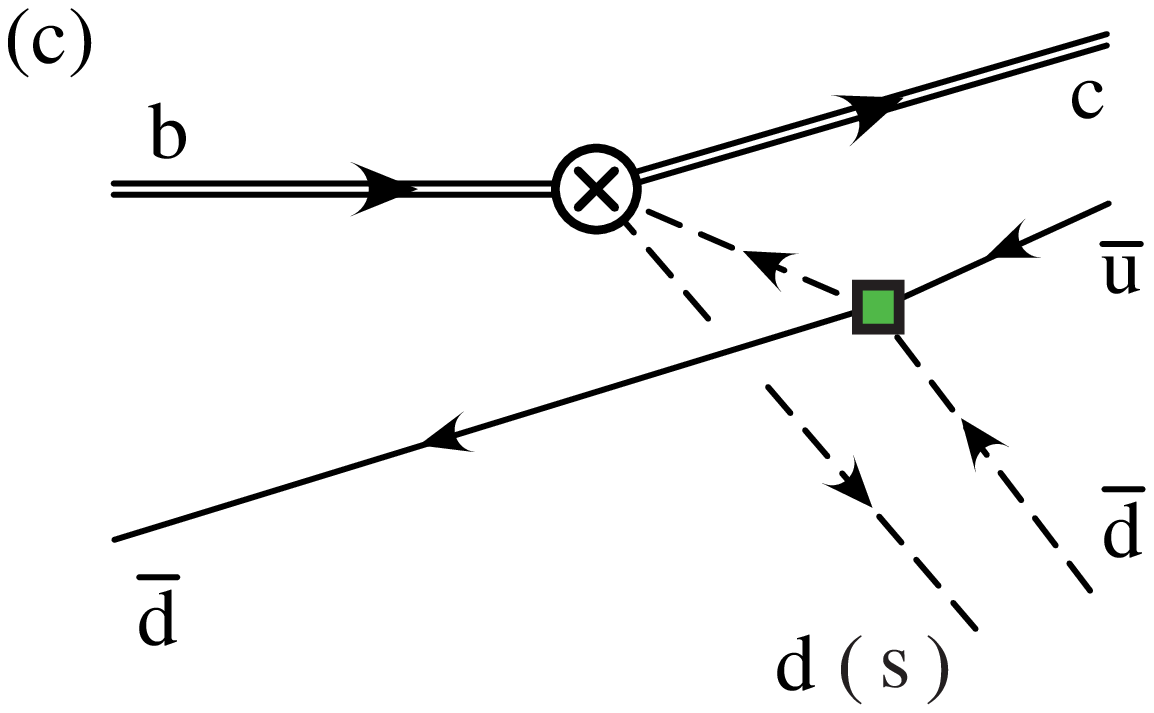}} }
  \hspace{0.cm}
  \raisebox{0.cm}{\mbox{\epsfxsize=5.7truecm
      \hbox{\epsfbox{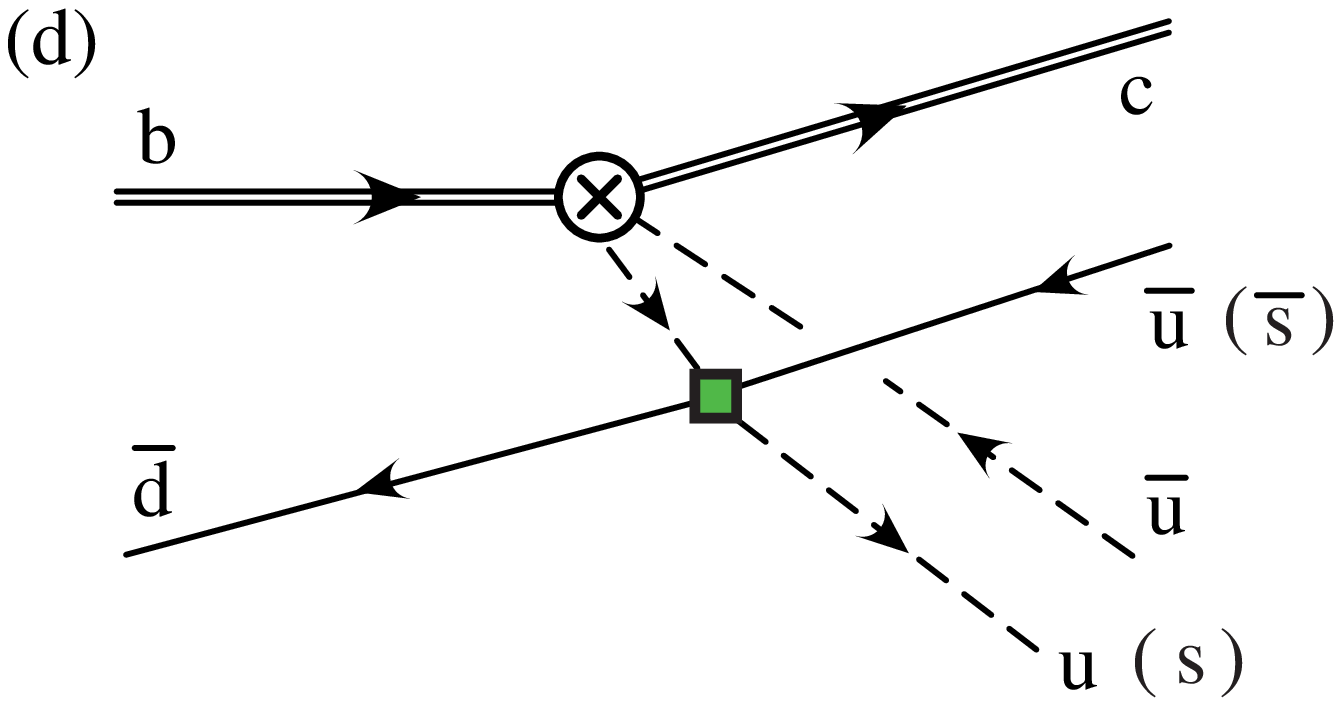}}} }
  \hspace{0.5cm}
  \raisebox{0.2cm}{\mbox{\epsfxsize=5.3truecm
      \hbox{\epsfbox{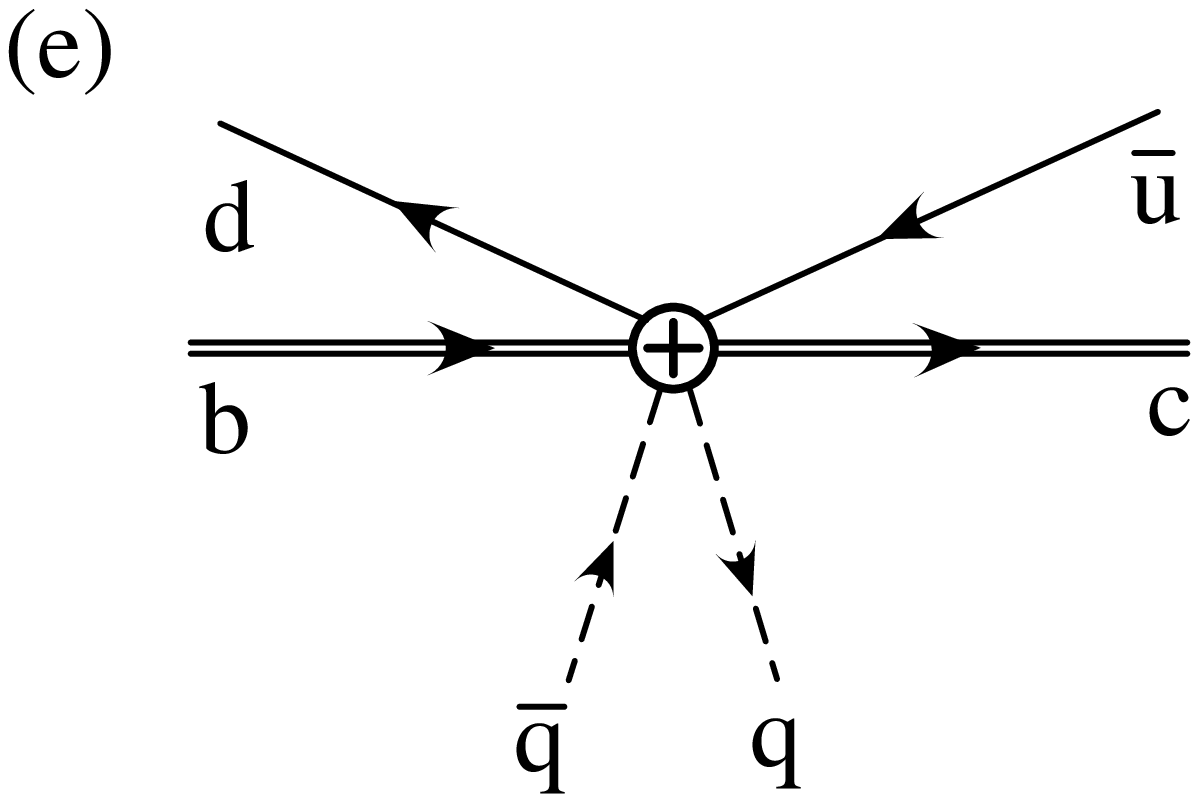}}} }
  }
\vskip-0.3cm
\caption[1]{Graphs for the tree level matching calculation from
  \SCETa (a,b) onto \SCETb (c,d,e) taken from Ref.~\cite{Mantry}. The dashed lines are collinear quark
  propagators and the spring with a line is a collinear gluon. Solid lines in
  (a,b) are ultrasoft and those in (c,d,e) are soft. The $\otimes$ denotes an
  insertion of the weak operator, given in Eq.~(\ref{QVI}) for (a,b) and in
  Eq.~(\ref{QV}) in (c,d). The $\oplus$ in (e) is a 6-quark operator from
  Eq.~(\ref{OV}).  The two solid dots in (a,b) denote insertions of the mixed
  usoft-collinear quark action ${\cal L}_{\xi q}^{(1)}$.  The boxes denote the
  \SCETb operator ${\cal L}_{\xi\xi qq}^{(1)}$~\cite{Mantry}.}
\label{fig_scet1} 
\vskip0cm
\end{figure}

For the color suppressed modes the $T$ topology does not contribute since the 
$W$ boson cannot produce the 
appropriate quark flavors needed to produce a neutral light meson at the weak
vertex. These modes receive contributions only from the $C$ and $E$
topologies where spectator quarks from the bottom or charmed mesons end up in
the pion. As in the case of the color allowed modes, after
the $W$ boson is integrated out, the weak vertex Hamiltonian $H_W$ is matched
onto the \SCETa operators ${\cal Q}^{(0,8)}_{j}(\tau_1,\tau_2)$. However, in
this case to produce a neutral collinear light meson and a charmed meson with
soft degrees of freedom, \SCETa requires a T-ordered product that involves two 
soft-hard-collinear transitions. 
\SCETa provides such a 
T-ordered product that involves two insertions of the subleading
ultrasoft-hard-collinear Lagrangian ${\cal L}_{\xi q}^{(1)}$:
\begin{eqnarray}\label{Tprod}
  T_j^{(0,8)} &=& \frac12 
  \int\!\! d^4\!x\, d^4\!y\ T\big\{ {\cal Q}_j^{(0,8)}(0)\,, 
   i{\cal L}_{\xi q}^{(1)}(x)\,, i{\cal L}_{\xi q}^{(1)}(y)\big\}\,.
\end{eqnarray}
Here the subleading Lagrangian is~\cite{bps4,Beneke:2002ph}
\begin{eqnarray}\label{Lxiq}
  {\cal L}^{(1)}_{\xi q} &=&   
 (\bar\xi_n W) \: \Big(\frac{1}{\bnP}\: W^\dagger ig\, \Bslash_\perp^c W\Big)
  q_{us} - \bar q_{us} \Big( W^\dagger ig\, \Bslash_\perp^c W 
  \frac{1}{\bnP^\dagger} \Big) (W^\dagger \xi_n) \,,
\end{eqnarray}
where $ig \Bslash_\perp^c =[i\bn\mcdot D^c,i\Dslash_\perp^{\,c}]$.
The two factors of $i{\cal L}_{\xi q}^{(1)}$ in Eq.~(\ref{Tprod}) are necessary to
swap one $u$ quark and one $d$ quark from ultrasoft to intermediate
collinear as shown in 
Fig.~\ref{fig_scet1}$(a,b)$.  In contrast 
to the $T$ topology, for this case both the ${\cal Q}^{(0)}_j$ and 
${\cal Q}^{(8)}_j$ color structures
can contribute. By power counting, the $T_j^{(0,8)}$'s are suppressed 
by $\lambda ^2 = \Lambda/Q$ relative to the leading operators for the $T$ topology. 

The \SCETa diagrams are now matched onto operators in \SCETb. These are shown in
Fig.~\ref{fig_scet1}$(b,c,d)$. In Figs.~\ref{fig_scet1}a,b
the gluon always has offshellness $p^2\sim E_M\Lambda$ due to momentum
conservation, and is shrunk to a point in \SCETb.  However, the collinear quark
propagator in (a,b) can either have $p^2\sim E_M\Lambda$ giving rise to the
short distance \SCETb contribution in Fig.~\ref{fig_scet1}e, or it can have
$p^2\sim \Lambda^2$ which gives the long distance \SCETb contribution in
Figs.~\ref{fig_scet1}c,d. It was shown~\cite{Mantry} to leading order in
$\alpha _s(Q)$ that the long distance contributions vanish for $M=\pi, \rho$.
For the kaons, these long distance contributions are non-vanishing but were shown to be
equal for $\bar B\to DK$ and $\bar B\to D^*K$ and for $\bar B\to DK^*_{||}$ and 
$\bar B\to D^*K^*_{||}$. In this section, we only review the analysis for $M=\pi
, \rho$ and refer the reader to Ref.~\cite{Mantry} for the kaon analysis.
Expressions for the long distance contributions are given in Ref.~\cite{Mantry}.
To all orders in
perturbation theory the Wilson coefficients $J^{(0,8)}$ from the matching of 
$\SCETa \to \SCETb$ generate only one spin structure and two color structures 
for the \SCETb short distance six quark operator:
\begin{eqnarray} \label{OV}
 O_{j}^{(0)}(k^+_i,\omega_k) &=& 
  \Big[ \bar h_{v'}^{(c)}  \Gamma^h_j\,  h_v^{(b)} \:
  (\bar d\,S)_{k^+_1} \nslash P_L\, (S^\dagger u)_{k^+_2} \Big]
  \Big[ (\bar \xi_n W)_{\omega_1} \Gamma_c (W^\dagger \xi_n)_{\omega_2} 
  \Big]\,, \\
 O_{j}^{(8)}(k^+_i,\omega_k) &=& 
  \Big[ (\bar h_{v'}^{(c)} S) \Gamma^h_j\, T^a\, (S^\dagger h_v^{(b)}) \:
  (\bar d\,S)_{k^+_1} \nslash P_L T^a (S^\dagger u)_{k^+_2} \Big]
  \Big[ (\bar \xi_n W)_{\omega_1} \Gamma_c (W^\dagger \xi_n)_{\omega_2} 
  \Big] \nn\,,
\end{eqnarray}
where the $d$, $u$, $h_{v'}^{(c)}$, and $h_v^{(b)}$ fields are soft, and
the $\xi_n$ fields are collinear isospin doublets, $(\xi_n^{(u)},\xi_n^{(d)})$.
In Eq.~(\ref{OV}) $\Gamma^h_{L,R}= \nslash P_{L,R}$ as in Eq.~(\ref{QVI}), while
for the collinear isospin triplet $\Gamma_c = \tau^3 \bnslash
P_L/2$. We do not list operators with a $T^a$ next to $\Gamma_c$ since
they will give vanishing contribution in the collinear matrix element. The
matrix elements of these operators gives the final result for the color
suppressed amplitude:

\begin{eqnarray}\label{result}
 A^{D^{(*)}}_{00} &=&  N_0^M
     \int_0^1\!\!\!dx\, dz\!\!  \int\!\! dk_1^+ dk_2^+\, 
   \Big[ C^{(i)}_{L}(z)\: 
   J^{(i)}(z,x,k_1^+,k_2^+)\: S^{(i)}_L(k_1^+,k_2^+)\:  \phi_M(x) \Big]\\
 &&\qquad\qquad\qquad\qquad\quad
   \pm  C^{(i)}_{R}(z)\: 
   J^{(i)}(z,x,k_1^+,k_2^+)\: S_R^{(i)}(k_1^+,k_2^+)\:  \phi_M(x) \Big] \nn \,,
\end{eqnarray}
where we sum over the color structures $i=0,8$ and $\pm$ refers to $D$ and $D^*$
respectively. The coefficients 
$C^{(0,8)}_{L,R}$ and $J^{(0,8)}$ are Wilson coefficients that arise in matching
calculations from $H_W\to \SCETa $ and $\SCETa \to\SCETb $ respectively. Tree
level expressions for $J^{(0,8)}$ are given in Eq.(42) of Ref.~\cite{Mantry}. The
non-perturbative functions $S_{L,R}$ are given by
\begin{eqnarray}\label{Sintro}
 \frac{\langle D^{0}(v') | (\bar h_{v'}^{(c)}S) \nslash P_{L,R} 
 (S^\dagger h_v^{(b)})
 (\bar d S)_{k^+_1}\nslash P_L (S^\dagger u)_{k^+_2} 
 | \bar B^0(v)\rangle}{\sqrt{m_B m_D}}
 &=& S_{L,R}^{(0)}(k^+_j) \,,\nn \\
\frac{ \langle D^{*0}(v',\varepsilon) | (\bar h_{v'}^{(c)}S) \nslash P_{L,R} 
 (S^\dagger h_v^{(b)}) (\bar d S)_{k^+_1} \nslash P_{L} (S^\dagger u)_{k^+_2} 
 | \bar B^0(v)\rangle}{\sqrt{m_B m_{D^*}}}
 &=&  \pm \frac{n\mcdot \varepsilon^*}{n\mcdot v'}\: S_{L,R}^{(0)}(k^+_j) \,.
\end{eqnarray}
Finally, the light cone wave functions are given by
[we suppress pre-factors of $\int_0^1 dx\, \delta(\omega_1-x\, \bn\mcdot
p_M)\, \delta(\omega_2+(1-x)\bn\mcdot p_M)$ on the RHS]\footnote{Our vector meson
  states are defined with an extra minus sign relative to the standard
  convention.}
\begin{eqnarray} \label{phipirho}
\langle \pi_n^0 |(\bar \xi_n W)_{\omega_1} \bnslash\gamma_5 \tau_3
  (W^\dagger \xi_n)_{\omega_2}| 0\rangle
  &=& -i\,\sqrt{2}\, f_\pi \: \bn\mcdot p_\pi\: \phi_\pi(\mu,x)\: \,,\\
\langle \rho_n^0(\varepsilon) |(\bar \xi_n W)_{\omega_1} \bnslash\tau_3
  (W^\dagger \xi_n)_{\omega_2}| 0\rangle
  &=& i\,\sqrt{2}\, f_\rho \: \bn\mcdot p_\rho\: \phi_\rho(\mu,x)\: \,. \nn
\end{eqnarray}
With $C_R=0$ at leading order in $\alpha _s(Q)$ in Eq.~(\ref{result}) we arrive
at the result in Eq.~(\ref{RM}) for the color suppressed modes.


\section{Excited Charmed Mesons} \label{sect_ECM}

Eqs.~(\ref{factLO}) and ~(\ref{result}) are the main results of the analysis for 
the $B\to D^{(*)}M$ decays. 
The analysis for decays  with excited charmed mesons $B\to
D^{**}M$ will proceed in exactly the same manner. Any difference in 
results  will show up only at the non-perturbative scale i.e. in \SCETb. 
In other words, the doublets $(D,D^*)$ and 
$(D_1,D_2^*)$ have the same quark content and any difference between them 
arises only
from non-perturbative effects responsible for their binding.
The physics at the scales $\mu ^2 \sim m_W^2$, $Q^2$, and $Q\Lambda
_{QCD}$ or in the theories $SM$, $H_W$, and $\SCETa $ is the same 
leaving the perturbative functions
$C_{1,2}$, $C^{(0,8)}_{L,R}$, and $J^{(0,8)}$ unchanged (see Fig. 3). The light 
cone wave function $\phi _{M}$  will also remain unchanged since the same final 
state light meson appears. At
leading order, the only change will be in the soft functions 
$S_{L,R}^{(i)}$ and $\xi $ since the matrix elements will now involve different 
non-perturbative final states namely $(D_1,D_2^*)$. We will denote the modified
functions as $Q_{L,R}^{(i)}$ and $\tau $ corresponding to
$S_{L,R}^{(i)}$ and $\xi $ respectively.

\subsection{Leading Order Predictions}

We now begin our analysis for the excited charmed states. We start by 
obtaining the modified soft functions $\tau $ and 
$Q_{L,R}^{(i)}$ and then carry over results for the perturbative 
functions and the non-perturbative collinear sector from the previous section to 
obtain the analog of Eqs.~(\ref{factLO}) and ~(\ref{result}). 

\subsubsection{Color Allowed Modes}

We first analyze the soft functions for the color allowed modes $\bar B^0\to
(D_1^+,D_2^{*+})M^-$ and $B^-\to
(D_1^0,D_2^{*0})M^-$. As before, the leading contribution to these modes 
comes from the $T$ topology which is given by the matrix elements of the
effective \SCETb operators ${\cal Q}^{(0,8)}_{L, R}$ of Eq.~(\ref{QV}). 
These matrix elements factorize into soft and collinear sectors. Using
the formalism of HQET, the soft part of the matrix element 
can be expressed in
general form as a trace

\begin{eqnarray}\label{traceca}
 \frac{\langle D_2^{*},D_1(v') | \bar h_{v'}^{(c)} \Gamma ^h_{L,R} h_v^{(b)} 
 | \bar B^0(v)\rangle}{\sqrt{m_B m_D}}
  =  \tau (\omega )\mbox{Tr } [v_{\sigma} \overline{F}_{v'}^{(c)\sigma } \Gamma H_v^{(b)}  ]\,,
\end{eqnarray}
where $\tau (\omega )$ is a new Isgur-Wise function analogous to 
$\xi (\omega )$. As in the case of ground state charmed mesons, 
the operators ${\cal Q}^{(8)}_{L, R}$ give vanishing contribution. 
$H_v^{(b)}$ and ${F}_{v'}^{(c)\sigma }$ in Eq.~(\ref{traceca}) are the superfields
for the heavy meson doublets $(\bar B, \bar B^*)$ and $(D_1,D_2^*)$ 
respectively~\cite{Falk:1990yz}
\begin{eqnarray}
  H_v &=& \frac{1+\vslash}{2}(P_v^{*\mu} \gamma_\mu + P_v \gamma_5)\nn \\
  F_v^{\sigma } &=& \frac{1+\vslash}{2}(D_2^{*\sigma \nu} \gamma_\nu -\sqrt \frac
  {3}{2}D_1^{\nu }\gamma_5[g^{\sigma }_{\nu } - \frac{1}{3} \gamma _{\nu }(\gamma
  ^{\sigma }-v^{\sigma })])\,.
\end{eqnarray}
As mentioned in the introduction, the matrix element in Eq.~(\ref{traceca}) which 
also appears in the case of semileptonic decays must vanish in the limit of 
zero recoil.
This condition is manifest in the right hand side of Eq.~(\ref{traceca}) 
through the property 
$v'_{\sigma} \overline{F}_{v'}^{(c)\sigma }=0$.
Thus, we expect the leading order amplitude to be proportional to some positive
power of $(\omega -1)$.
At maximum recoil $(\omega _0-1)\sim 0.3\sim \Lambda _{QCD}/Q$ putting 
$(\omega _0-1)$ and
$\Lambda _{QCD}/Q$ on the same footing in the power counting scheme. 
In addition, maximum recoil is a special kinematic point where the heavy meson 
masses are 
related to $\omega _0$  through $(\omega _0 -1)=\frac {(m_B-m_D)^2}{2m_Bm_D}$. 
We must keep this relation in mind to make the power counting manifest and
so it becomes convenient to express $(m_B-m_D)$ in terms of $(\omega _0-1)$.

Computing the trace in Eq.~(\ref{traceca}) we arrive at the result 
\begin{eqnarray}
\frac{\langle D_1(v') | \bar h_{v'}^{(c)} \Gamma ^h_{L,R}h_v^{(b)} 
 | \bar B^0(v)\rangle}{\sqrt{m_B m_D}} &=&  \tau (\omega )\sqrt {\frac
 {m_B(\omega +1)}{3m_D}}\epsilon
^*\cdot v \nn \\
\frac{\langle D_2^{*}(v') | \bar h_{v'}^{(c)} \Gamma ^h_{L,R}h_v^{(b)} 
 | \bar B^0(v)\rangle}{\sqrt{m_B m_D}} &=& \pm \tau (\omega )\sqrt{\frac {m_B}{2m_D(\omega -1)}}
 \epsilon ^{*\sigma
\nu}v_\sigma v_\nu ,
\end{eqnarray}
where the $\pm $ for the $D_2^*$ refer to the choice of $\Gamma ^h_L$ and 
$\Gamma ^h_R$ Dirac structures respectively. 
$\epsilon ^\mu $ and $\epsilon ^{\mu \nu}$ are the 
polarizations for $D_1$ and $D_2^*$ respectively.
Combining this result for the soft sector with the hard and collinear parts from
the previous section we obtain the final result 
\begin{eqnarray}\label{caamp} 
  A(B\to D_1M) &=& N^{D1}E_M\sqrt {\frac
 {m_B(\omega _0 +1)}{3m_D}}\epsilon
^*\cdot v \: \tau (w_0,\mu) \nn \\
&\times & \int_0^1\!\!dx\: 
   T^{D_1}(x,m_c/m_b,\mu)\: \phi_M(x,\mu)  \nn \\
   A(B\to D_2^*M) &=& N^{D2*}E_M\sqrt{\frac {m_B}{2m_D(\omega _0 -1)}}
 \epsilon ^{*\sigma
\nu}v_\sigma v_\nu \: \tau (w_0,\mu) \nn \\
&\times &\int_0^1\!\!dx\: 
   T^{D_2^*}(x,m_c/m_b,\mu)\: \phi_M(x,\mu)\,,
\end{eqnarray}
where the normalizations are given by
\begin{eqnarray}\label{norm1} 
N^{D1}=\frac {G_FV_{cb}V^*_{ud}}{\sqrt 2} f_M\sqrt
{m_Bm_{D1}}, \qquad
N^{D2*}=\frac {G_FV_{cb}V^*_{ud}}{\sqrt 2} f_M\sqrt
{m_Bm_{D2*}}
\end{eqnarray}
and the hard kernels $T^{(D_1,D_2^*)}(x,\mu)$ are the same as those appearing
in Eq.~(\ref{factLO}) $T^{(D_1,D_2^*)}(x,\mu)=T^{(*)}(x,\mu)$. Using the properties
of the polarization sums
\begin{eqnarray}\label{polarizationsum}
\sum _{pol}|\epsilon ^{*}\cdot v|^2=(\omega +1)(\omega -1)\,,\qquad
\sum _{pol}|\epsilon ^{*\sigma \nu}v_\sigma v_\nu|^2=\frac {2}{3}(\omega
+1)^2(\omega -1)^2,
\end{eqnarray}
the unpolarized amplitude squared is given by
\begin{eqnarray}\label{normsum}
\sum _{pol}|A(B\to (D_1,D_2^*)M)|^2&=&|N^{(D1,D2*)}\int_0^1\!\!dx\: 
   T^{(D1,D_2^*)}(x,m_c/m_b,\mu)\: \phi_M(x,\mu)|^2 \nn \\
   &\times &\frac {m_B\tau ^2(\omega ,\mu )}{3m_D}(\omega _0+1)^2(\omega _0-1).
\end{eqnarray}
At leading order in $\Lambda _{QCD}/m_{b,c}$ the masses in the heavy quark 
doublet $(D_1,D_2^*)$ are degenerate giving the relation $N^{(D_1)}=N^{(D_2*)}$.
In addition at leading order in $\alpha _s(Q)$, $T^{(D_1)}=T^{(D_2*)}$ allowing
us to make a prediction for the unpolarized color allowed branching ratios:
\begin{eqnarray}\label{brpred1}
  \frac{Br(\bar B^0 \to D_2^{*+}M^-)}
    {Br(\bar B^0 \to D_1^+M^-)} = \frac {Br(B^- \to D_2^{*0}M^-)}
    {Br(B^- \to D_1^0M^-)}=1 \,.
\end{eqnarray} 
The same result was derived in ref.~\cite{Leibovich:1997em} at 
lowest order in $1/m_{b,c}$ by evaluating their results for semileptonic decays at the maximum recoil 
point and replacing the $e\bar \nu$ pair with a massless pion. Recently, a 
theoretical prediction of 0.91 for the above ratio was made in the covariant 
light front model~\cite{Cheng:2003sm}.

\subsubsection{Color Suppressed Modes}

Now we look at the color suppressed modes $\bar
B^0\to (D_1^0,D_2^{*0})M^0$. The leading contributions are 
from the $C$ and $E$ topologies which are given by matrix elements of the \SCETb
operators $O_{j}^{(0,8)}(k^+_i,\omega_k)$ of 
Eq.~(\ref{OV}). Once again, the result factorizes and using the formalism of HQET,
the soft part of the matrix element can be expressed as a trace

\begin{eqnarray}\label{trace}
 \frac{\langle D_2^{(*)0},D_1^0(v') | (\bar h_{v'}^{(c)} S) \Gamma _j^h\, 
  (S^\dagger h_v^{(b)}) \:(\bar d\,S)_{k^+_1} \nslash P_L\, 
  (S^\dagger u)_{k^+_2}
 | \bar B^0(v)\rangle}{\sqrt{m_B m_D}}
  =  \mbox{Tr } [\overline{F}_{v'}^{(c)\sigma } \Gamma _j^hH_v^{(b)} X^{(0)}_{\sigma
  } ]\,,
\end{eqnarray} 
with similar expressions for the $O_{j}^{(8)}(k^+_i,\omega_k)$ operators.
The Dirac structure $X^{(0,8)}_{\sigma }$ is of the most general form allowed by
the symmetries of QCD and involves eight form factors
\begin{eqnarray}\label{genformcs}
X^{(0,8)}_{\sigma }=v_{\sigma }(a_1^{(0,8)}\nslash P_L+a_2^{(0,8)}\nslash P_R+
a_3^{(0,8)}P_L+a_4^{(0,8)} P_R) \nonumber \\
+ n_{\sigma }(a_5^{(0,8)}\nslash P_L+
a_6^{(0,8)}\nslash P_R+
a_7^{(0,8)}P_L+a_8^{(0,8)} P_R)
\end{eqnarray}
Computing the trace in Eq.~(\ref{trace}), the soft matrix elements are given by
\begin{eqnarray}\label{softmelt}
\frac{\langle D_2^{*0}(v') | (\bar h_{v'}^{(c)} S) \Gamma ^h_{L,R}\, 
  (S^\dagger h_v^{(b)}) \:(\bar d\,S)_{k^+_1} \nslash P_L\, (S^\dagger u)_{k^+_2}
 | \bar B^0(v)\rangle}{\sqrt{m_B m_D}}
 &=&\frac {(\mp \epsilon ^{*\sigma \nu}v_{\sigma }v_{\nu
})Q^{(0)}_{L,R}}{4(\omega +1)(\omega -1)} \nonumber \\
\frac{\langle D_1^0(v') | (\bar h_{v'}^{(c)} S) \Gamma ^h_{L,R}\, 
  (S^\dagger h_v^{(b)}) \:(\bar d\,S)_{k^+_1} \nslash P_L\, (S^\dagger u)_{k^+_2}
 | \bar B^0(v)\rangle}{\sqrt{m_B m_D}}&=&
 \frac{(\epsilon ^*\cdot v)Q^{(0)}_{L,R}}{
 \sqrt {24(\omega +1)(\omega -1)}}
 \end{eqnarray}
where,
\begin{eqnarray}\label{hatsoft}
Q^{(0)}_L&=& \frac {-1}{m_B^2m_D^2}[2m_Bm_D(2a_1^{(0)}m_B^2-a_3^{(0)}m_B^2-
a_4^{(0)}m_Bm_D)\sqrt {(\omega +1)(\omega -1)}\nonumber \\
&+&4a_5^{(0)}m_B^4-2a_7^{(0)}m_B^4-2a_8^{(0)}m_Dm_B^3]\nonumber \\
Q^{(0)}_R&=&\frac {-1}{m_B^2m_D^2}[2m_Bm_D(2a_2^{(0)}m_B^2-a_3^{(0)}m_Bm_D-
a_4^{(0)}m_B^2)\sqrt {(\omega +1)(\omega -1)} \nonumber \\
&+&4a_6^{(0)}m_B^4-2a_7^{(0)}m_Dm_B^3-2a_8^{(0)}m_B^4],
\end{eqnarray}
with similar expressions for $Q_{L,R}^{(8)}$. Here the soft functions
$Q_{L,R}^{(0,8)}$ are the analog of $S_{L,R}^{(0,8)}$ in Eq.~(\ref{result}). 
It was shown~\cite{Mantry} that these soft functions generate a
non-perturbative strong phase. We note that in both the $D_1$ and $D_2^*$
decay channels, since the same moments of the non-perturbative functions $Q_{L,R}^{(0,8)}$ appear, 
their strong phases are predicted to be equal
\begin{eqnarray}\label{strongphi}
\phi _{D_1M}=\phi _{D_2^*M}. 
\end{eqnarray}
The analogous strong phase $\phi$ for $\bar B^0\to D^{(*)0}\pi ^0$ is shown in Fig.~\ref{fig_isospin}.
Since the strong phases $\phi$ and $\phi _{D_1\pi,D_2^*\pi}$ are determined by different 
non-perturbative functions $S_{L,R}^{(0,8)}$ and
$Q_{L,R}^{(0,8)}$ respectively, we do not expect them to be related.

Keeping in mind that the perturbative functions $C_{L,R}^{(i)}$ and $J^{(i)}$ remain 
unchanged, we can combine the result in Eq.~(\ref{softmelt}) for soft sector with the 
collinear and hard parts of the amplitude to arrive at the result
\begin{eqnarray}\label{result2}
 A^{(D1)}_{00} &=&  \frac{-N^{D_1}\epsilon ^*\cdot v}{\sqrt {24(\omega _0+1)(\omega _0-1)}}
     \int_0^1\!\!\!dx\, dz\!\!  \int\!\! dk_1^+ dk_2^+\, 
    \Big[ C^{(i)}_{L}(z)\: 
   J^{(i)}(z,x,k_1^+,k_2^+)\: Q^{(i)}_L(k_1^+,k_2^+)\:  \phi_M(x) \nn \\
  &&
    - C^{(i)}_{R}(z)\: 
   J^{(i)}(z,x,k_1^+,k_2^+)\: Q_R^{(i)}(k_1^+,k_2^+)\:  \phi_M(x) \Big]\nn \\ 
 A^{(D2*)}_{00} &=&  \frac {N^{D_2^*}\epsilon ^{*\sigma \nu}v_{\sigma }v_{\nu
}}{4(\omega _0+1)(\omega _0-1)}
     \int_0^1\!\!\!dx\, dz\!\!  \int\!\! dk_1^+ dk_2^+\, 
    \Big[  C^{(i)}_{L}(z)\: 
   J^{(i)}(z,x,k_1^+,k_2^+)\: Q^{(i)}_L(k_1^+,k_2^+)\:  \phi_M(x) \nn \\
  &&
    + C^{(i)}_{R}(z)\: 
   J^{(i)}(z,x,k_1^+,k_2^+)\: Q_R^{(i)}(k_1^+,k_2^+)\:  \phi_M(x) \Big] 
   \, .
\end{eqnarray}
Once again the vanishing of $C_R^{(0,8)}$ in Eq.~(\ref{result2}) at leading order 
in $\alpha _s(Q)$ and using the polarization sums in Eq.~(\ref{polarizationsum})
gives the unpolarized amplitude squared
\begin{eqnarray}\label{unpolcs}
\sum _{pol}|A_{00}^{(D_1,D_2*)}|^2&=&\frac {1}{24}\Big|N^{(D_1,D_2^*)}\int_0^1\!\!\!dx\, dz\!\!  \int\!\! dk_1^+ dk_2^+\, 
    \Big[  C^{(i)}_{L}(z)\: 
   J^{(i)}(z,x,k_1^+,k_2^+)\: \nn \\
   &\times &Q^{(i)}_L(k_1^+,k_2^+)\:  \phi_M(x)\Big]\Big|^2  .
\end{eqnarray}
Since $N^{D_1}=N^{D_2*}$, at leading order in $\Lambda _{QCD}/m_Q$ we can
make a prediction for the unpolarized branching ratios
\begin{eqnarray}\label{excsratio}
\frac {Br(\bar B^0 \to D_2^{*0}M^0)}{Br(\bar B^0 \to D_1^{0}M^0)}=1,
\end{eqnarray}
which is one of the main results of this paper. Note that from the point of
view of naive factorization, this result is quite unexpected since the tensor
meson $D_2^*$ cannot be produced by a V-A current.



\subsection{Power Counting and Next to Leading Order Contributions}\label{sect_NLO}

\subsubsection{Color Allowed Modes}

We see that as required by HQS, the unpolarized amplitude in Eq.~(\ref{normsum})
is proportional to $(\omega _0-1)$ which is expected to provide a suppression
of this leading order result. However, it is also accompanied by a
factor of $(\omega _0 +1)^2$. At maximum recoil $\omega _0$
is related to the energy of the light meson and the mass of the charmed meson
through
\begin{eqnarray}\label{maxkinematic}
\sqrt{(\omega _0+1)(\omega _0-1)}=\frac {E_M}{m_D}.
\end{eqnarray}
Thus, in the SCET power counting scheme the quantity 
$\sqrt{(\omega _0+1)(\omega _0-1)}$ is of order one. It is now clear from 
Eq.~(\ref{normsum}) and the above relation that despite the constraint of HQS
there is no suppression of the leading order result and
the subleading corrections of order $\Lambda _{QCD}/Q$ are not dangerous to the
leading order result. This allows us to rely on the leading order predictions 
up to corrections
supressed by $\Lambda _{QCD}/Q$.

\begin{figure}[h]
  \epsfxsize = 10cm
  \centerline{\epsfbox{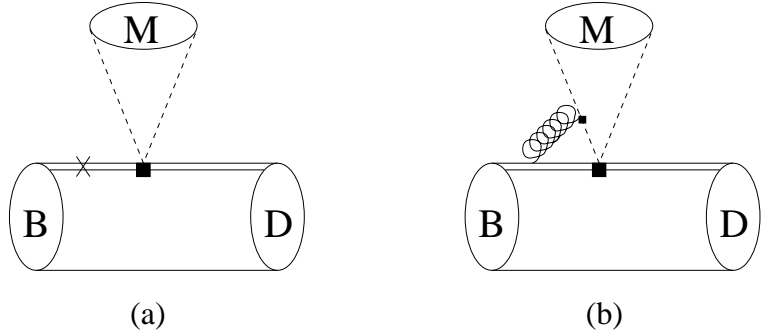}}
  \caption{Contributions to the color allowed sector from T-ordered products of
  the effective weak vertex in \SCETa with subleading kinetic and chromomagnetic
  HQET operators (a) and with the subleading SCET operators(b). In this section, to illustrate through examples the relative 
  suppression the subleading contributions by at least $\Lambda _{QCD}/Q$, we
  only consider T-ordered products of type (a). The analysis for type (b)
  contributions will proceed in a similar manner. }
  \label{tproduct}
\end{figure}

To illustrate the above ideas, in this section we will compute some of the
subleading corrections and compare their sizes relative to the leading order
predictions. The leading order operators in Eqs.~(\ref{QV}) and ~(\ref{OV})
are products of soft and collinear operators $O=O_s*O_c$. Subleading 
corrections can arise in four possible ways
\begin{itemize}
\item corrections in the soft sector to $O_s$ and from T-products(see
Fig.~\ref{tproduct}a) with $O_s$. 
\item corrections in the collinear sector to $O_c$ and from 
T-products(see for example Fig.~\ref{tproduct}b) with $O_c$.
\item corrections from subleading mixed collinear-soft  operators and 
their T-products.
\item Beyond the heavy quark limit, $s_l$ is no longer a good quantum number.
From table~\ref{table_hqs}, we see that it implies mixing between $D_1$ and $D_1^*$.
Thus, the physical $D_1$ state will have a small admixture of the $D_1^*$ state beyond
the heavy quark limit which will play a role in subleading corrections.
\end{itemize}
We will only focus on subleading corrections in the soft sector 
from HQET as in Fig.~\ref{tproduct}a in order to
illustrate the power counting. These corrections give precisely the subleading 
semileptonic form factors which were computed in Ref.~\cite{Leibovich:1997em}. 
The analysis for the remaining subleading corrections will follow
in a similar manner and we leave it as possible future work.

The HQET and QCD fields are related to eachother through

\begin{eqnarray}\label{QCDHQET}
Q(x)=e^{-im_Qv\cdot x}[1+\frac{i\Dslash }{2m_Q}+\dots ]h_v^{(Q)},
\end{eqnarray}
where the ellipses denote terms suppressed by higher orders of 
$\Lambda _{QCD}/m_Q$ and $Q=b,c$. Including the $\Lambda _{QCD}/m_Q$ corrections,
the QCD current is now matched onto

\begin{eqnarray}\label{NLOcurrent}
\bar c\Gamma b \to \bar h^{(c)}_{v'}(\Gamma -\frac {i}{2m_c}\lvec \Dslash \Gamma + 
\frac {i}{2m_b}\Gamma \vec \Dslash )h_v^{(b)}.
\end{eqnarray}
Then there are subleading corrections from T-ordered products of the leading
order current with order $\Lambda _{QCD}/Q$ terms in the HQET Lagrangian:
\begin{eqnarray}\label{LHQET}
\delta L_{HQET}= \frac {1}{2m_Q}[O^{(Q)}_{kin,v}+O^{(Q)}_{mag,v}]
\end{eqnarray}
where $O^{(Q)}_{kin,v}$ and $O^{(Q)}_{mag,v}$ are the kinetic and chromomagnetic
operators

\begin{eqnarray}\label{kin_mag}
O^{(Q)}_{kin,v}=\bar h^{(Q)}_{v}(iD)^2h_v^{(Q)} , \qquad
O^{(Q)}_{mag,v}=\bar h^{(Q)}_{v}\frac {g_s}{2}\sigma _{\alpha \beta
}G^{\alpha \beta }h_v^{(Q)} .
\end{eqnarray}
We employ the trace formalism to compute these subleading corrections to the 
soft matrix element from corrections to the matching in Eq.~(\ref{NLOcurrent})
\begin{eqnarray}\label{NLOfields}
\bar h^{(c)}_{v'}i\lvec D_{\lambda }\gamma ^{\lambda }\Gamma
h_v^{(b)}&=&Tr[S_{\sigma \lambda }^{(c)}\bar F^{\sigma (c)}_{v'}\gamma
^{\lambda }\Gamma H_v^{(b)}] \nn \\
\bar h^{(c)}_{v'}\Gamma \gamma ^{\lambda }i\vec D_{\lambda }
h_v^{(b)}&=&Tr[S_{\sigma \lambda }^{(b)}\bar F^{\sigma (c)}_{v'}\Gamma \gamma
^{\lambda }H_v^{(b)}],
\end{eqnarray}
and from T-ordered products with $\delta L_{HQET}$
\begin{eqnarray}\label{TOmag}
i\int d^4x T(O^{(c)}_{mag,v'}(x)[\bar h^{(c)}_{v'}\Gamma h_v^{(b)}](0)) &=&
Tr[R_{\sigma \alpha \beta }^{(c)}\bar F^{\sigma (c)}_{v'}i\sigma ^{\alpha \beta
}\frac {1+\vslash '}{2}\Gamma H_v^{(b)}] 
 \nn \\
i\int d^4x T(O^{(b)}_{mag,v}(x)[\bar h^{(c)}_{v'}\Gamma h_v^{(b)}](0)) &=&
Tr[R_{\sigma \alpha \beta }^{(b)}\bar F^{\sigma (c)}_{v'}
\Gamma \frac {1+\vslash }{2}i\sigma ^{\alpha \beta
} H_v^{(b)}] 
\end{eqnarray}
where the structures $S_{\sigma \lambda}^{(Q)}$ and $R^{(Q)}$ 
are parametrized as
\begin{eqnarray}\label{Sformfactor}
S_{\sigma \lambda }^{(Q)}&=&v_{\sigma }[\tau _1^{(Q)}v_\lambda +
\tau_2^{(Q)}v'_\lambda + \tau ^{(Q)}_3\gamma _{\lambda }] + \tau
^{(Q)}_4g_{\sigma \lambda } \nn \\
R_{\sigma \alpha \beta }^{(c)}&=&n_1^{(c)}v_{\sigma }\gamma _{\alpha }\gamma
_{\beta }+n_2^{(c)}v_{\sigma }v _{\alpha }\gamma
_{\beta }+n_3^{(c)}g_{\sigma \alpha }v
_{\beta } ,\nn \\
R_{\sigma \alpha \beta }^{(b)}&=&n_1^{(b)}v_{\sigma }\gamma _{\alpha }\gamma
_{\beta }+n_2^{(b)}v_{\sigma }v' _{\alpha }\gamma
_{\beta }+n_3^{(b)}g_{\sigma \alpha }v'
_{\beta } .
\end{eqnarray}
The T-ordered products with the kinetic energy operator $O_{kin,v}^{(Q)}$
do not violate spin symmetry and simply provide $\Lambda _{QCD}/m_Q$
corrections to the form factor in Eq.~(\ref{traceca}) 
$\tau \to \tilde \tau = \tau + \frac {\eta _{ke}^{c}}{2m_c}
+ \frac {\eta _{ke}^{b}}{2m_b}$. 
The form factors appearing in $S_{\sigma \lambda }^{(Q)}$ are not all
independent and are related~\cite{Leibovich:1997em} through
\begin{eqnarray}\label{tauconstraint}
\omega \tau_1^{(c)}+\tau_2^{(c)}-\tau _3^{(c)}&=&0 \nn \\
\tau_1^{(b)}+\omega \tau_2^{(b)}-\tau _3^{(b)}+\tau _4^{(b)}&=&0 \nn \\
\tau _1^{(c)}+\tau _1^{(b)}&=&\bar \Lambda \tau  \nn \\
\tau _2^{(c)}+\tau _2^{(b)}&=&-\bar \Lambda '\tau  \nn \\
\tau _3^{(c)}+\tau _3^{(b)}&=&0 \nn \\
\tau _4^{(c)}+\tau _4^{(b)}&=&0 ,
\end{eqnarray}
where $\bar \Lambda $ and $\bar \Lambda '$ are the energies of the light degrees of
freedom in the $m_{b,c}\to \infty $ limit for the $(\bar B,\bar B^*)$ and  
$(D_1,D_2^*)$ HQS doublets respectively.
Using these relations we can express our results in terms of the
$\tau $, $\tau _1^{(c)},$ and $\tau _2^{(c)}$ form factors. 
Combining the subleading contributions from Eqs.~(\ref{NLOfields}) and
~(\ref{TOmag}) with the leading order result in
Eq.~(\ref{traceca}) and using constraints from Eq.~(\ref{tauconstraint}) 
we can write the soft matrix element as
\begin{eqnarray}
S_{D1,D2*}&=&
\tilde \tau (\omega _0)\mbox{Tr } [v_{\sigma} \overline{F}_{v'}^{(c)\sigma } \Gamma
H_v^{(b)}  ]-\frac {1}{2m_c}Tr[S_{\sigma \lambda }^{(c)}\bar F^{\sigma (c)}_{v'}\gamma
^{\lambda }\Gamma H_v^{(b)}]-\frac {1}{2m_b}Tr[S_{\sigma \lambda }^{(c)}\bar F^{\sigma (c)}_{v'}\Gamma \gamma
^{\lambda }H_v^{(b)}] \nn \\
&+&\frac {\tau (\omega )}{2m_b}Tr[(\bar \Lambda v_\lambda -
\bar \Lambda 'v'_{\lambda })v_{\sigma }\bar F^{\sigma (c)}_{v'}\Gamma \gamma
^{\lambda }H_v^{(b)}]+\frac {1}{2m_c}Tr[R_{\sigma \alpha \beta }^{(c)}\bar F^{\sigma (c)}_{v'}i\sigma ^{\alpha \beta
}\frac {1+\vslash '}{2}\Gamma H_v^{(b)}] \nn \\
&+&\frac {1}{2m_b}Tr[R_{\sigma \alpha \beta }^{(b)}\bar F^{\sigma (c)}_{v'}
\Gamma \frac {1+\vslash }{2}i\sigma ^{\alpha \beta
} H_v^{(b)}]+\cdots ,
\end{eqnarray}
where the ellipses denote contributions from other subleading operators that we
have not considered. Computing the above traces and combining the results for
the hard and collinear parts from section II,
the amplitudes can be brought into the final form
\begin{eqnarray}
A(B\to D_1M)=N^{D1}f^{BD1}\epsilon ^*\cdot v
\: \int_0^1\!\!dx\: 
   T^{D_1}(x,m_c/m_b,\mu)\: \phi_M(x,\mu)  \nn \\
A(B\to D_2^*M)=N^{D2*}f^{BD2*}\epsilon ^{*\sigma \nu }v_\sigma v_\nu
\: \int_0^1\!\!dx\: 
   T^{D_2*}(x,m_c/m_b,\mu)\: \phi_M(x,\mu),    
\end{eqnarray}
where $f^{(BD1,BD2*)}$ are functions of the form factors
$\tilde \tau ,\tau _1^{(c)},\tau _2^{(c)}, \eta
_{1,2,3}^{(c,b)}$. For the $D_1$ channel, $f^{BD1}$
is given by
\begin{eqnarray}\label{fBD1}
\sum _{pol}|\epsilon ^{*}\cdot v\bar f^{BD1}|^2&=&\frac {m_B(\omega _0
+1)}{12m_D} \nn \\
&\times &\Big[\Big(2\tilde \tau +\frac {6\eta _{1}^{(c)}}{m_c}
-\frac
{2\eta _{2}^{(c)}}{m_c}-\frac {\eta _{3}^{(c)}}{m_c} 
+\frac {6\eta _{1}^{(b)}}{m_b}+\frac
{\eta _{2}^{(b)}}{m_b}+\frac {\eta _{3}^{(b)}}{m_b} \nn \\
&+&(\frac {m_B}{m_D}
+\frac {m_D}{m_B})\frac {\eta
_2^{(c)}}{m_c} -(\frac {m_B}{m_D}+\frac {m_D}{m_B}-1)\frac {\eta _2^{(b)}}{m_b}
\Big)\sqrt {(\omega _0+1)(\omega _0-1)}  \\
 &-&\big(\frac {m_B^2}{m_D^2}+\frac {m_D^2}{m_B^2}\big)\frac {\tau _1^{(c)}}{2m_c}
+\big(\frac {\tau _2^{(c)}}{m_c}-\frac {2\bar \Lambda \tau}{m_c}\big)
+\big(\frac {m_B}{m_D}+\frac {m_D}{m_B}\big)
\big(\frac {\bar \Lambda '\tau}{m_c} +\frac {\tau
_1^{(c)}}{2m_c}-\frac {\tau
_2^{(c)}}{2m_c}\big) \nn \\
&+&\big(\frac {m_B}{m_D}+\frac {m_D}{m_B}+1\big)\frac {\tau _1^{(c)}}{m_b}\sqrt
{(\omega _0
-1)}
+\big(\frac {\tau _2^{(c)}}{m_b}-\frac {\bar \Lambda \tau
}{m_b}-\frac {\bar \Lambda '\tau }{m_b}\big)(\omega _0-1)+\cdots \Big ]^2, \nn
\end{eqnarray}
and for the $D_2$ channel, $f^{BD2}$ is given by
\begin{eqnarray}\label{fBD2*}
\sum _{pol}|\epsilon ^{*\sigma \nu }v_\sigma v_\nu\bar f^{BD2*}|^2&=&
\frac {m_B(\omega _0+1)}{12m_D}\nn \\
&\times &\Big[\Big(2\tilde \tau -\frac {2\eta _1^{(c)}}{m_c}+ \frac {\eta
_2^{(c)}}{m_c} 
+\frac {\eta _3^{(c)}}{m_c}-\frac {\tau
_{2}^{(c)}}{m_c} 
+\frac {6\eta _1^{(b)}}{m_b}+\frac {\eta _2^{(b)}}{m_b}
+\frac {\eta _3^{(b)}}{m_b} \nn \\
&+&(\frac {m_B}{m_D}+\frac {m_D}{m_B}-1)(\frac{\tau
_1^{(c)}}{m_c}-\frac{\eta
_2^{(b)}}{m_b})\Big)\sqrt {(\omega _0+1)(\omega _0-1)} \\
&-& \Big((\bar
\Lambda +\bar \Lambda ')\frac {\tau }{m_b}-\frac {\tau _2^{(c)}}{m_b}-(\frac
{m_B}{m_D}+\frac {m_D}{m_B}+1)(\frac {\eta _2^{(c)}}{m_c}-\frac {\tau
_1^{(c)}}{m_b})\Big)
(\omega _0 -1)+\cdots \Big]^2. \nn
\end{eqnarray}
The above expressions are written in a way to make the power counting manifest.
The ratio $\frac {m_B}{m_D}$ is of order one, $(\omega _o-1)$ is numerically of
order $\Lambda _{QCD}/Q$,
and as discussed in Eq.~(\ref{maxkinematic}) the quantity 
$\sqrt {(\omega _0-1)(\omega _0+1)}$ is of 
order one. We see that 
the leading order contribution inside the square brackets in Eqs.~(\ref{fBD1}) 
and ~(\ref{fBD2*}) is proportional to 
$\tau \sqrt {(\omega _0 -1)(\omega _0+1)}$ and is the same for the $D_1$ and
$D_2^*$ 
channels. More importantly, there is no suppression of the leading
order term due to HQS since $\sqrt {(\omega _0 -1)(\omega _0+1)}$ is of 
order
one. On the other hand, the subleading corrections in the
square brackets are of size either $\Lambda _{QCD}/m_Q$, $(\omega _0 -1)\Lambda
_{QCD}/m_Q$, $\sqrt {(\omega _0-1)}\Lambda
_{QCD}/m_Q$, or $\sqrt {(\omega _0-1)(\omega _0+1)}\Lambda
_{QCD}/m_Q$ and hence are suppressed by at least
$\Lambda _{QCD}/m_Q$ relative to the leading order prediction. Thus, we see that
the constraints of HQS enter in a very specific manner so as to preserve
the power counting scheme of SCET allowing us to
ignore the subleading corrections near maximum recoil. It was the maximum recoil relation in
Eq.~(\ref{maxkinematic}) that ensured no suppression of the
leading order result.
The predictions of Eq.~(\ref{brpred1}) remain intact with these subleading corrections 
suppressed by at least $\Lambda _{QCD}/Q$.

\subsubsection{Color Suppressed Modes}

In the case of color suppressed decays which are mediated by operators 
that are not conserved currents, there is no reason to expect the soft matrix
element to vanish at zero recoil by HQS and thus no reason to expect a
suppression at maximum recoil. 
In fact the  
non-trivial dependence of the soft matrix elements in Eq.~(\ref{trace})
on the light cone vector $n_\mu $ makes it difficult to make a comparison with 
the zero recoil limit. The soft functions $Q_{L,R}^{(0,8)}$  will
depend on the light cone
vector $n^\mu $ through the arguments
$(n\cdot v, n\cdot v', n\cdot k_1, n\cdot k_2)$ and it is not obvious how to
extrapolate such a function away from maximum recoil. At maximum recoil
$v$, $v'$ and $n$ are related through
\begin{eqnarray}\label{kin}
m_Bv^\mu = m_Dv^{'\mu }+E_Mn^\mu .
\end{eqnarray} 
The light cone vector has
the special property $n^2=0$ and is a reflection of the onshell condition of 
the pion $p^2_\pi =(E_\pi n)^2=0$. Away from maximum recoil, $E_Mn^\mu$ is to
be
replaced by $q^\mu$ which is offshell $q^2\neq 0$, 
inconsistent with the $n^2=0$ property of the light cone vector. So, 
Eq.~(\ref{kin}) can no longer be used to determine $n^\mu$ in terms of $v^\mu$ and
$v^{'\mu}$ and thus more care is required in extrapolating away from maximum 
recoil. 
  
From Eqs.~(\ref{result2}), ~(\ref{hatsoft}), and ~(\ref{maxkinematic}) and the
power counting scheme discussed earlier we see 
that there is in fact no suppression of the leading order color suppressed
amplitude. The leading order predictions of
Eq.~(\ref{excsratio}) remain intact with corrections suppressed by at least 
$\Lambda _{QCD}/Q$. We leave the analysis of subleading corrections in the 
color suppressed sector as possible future work.


\section{Phenomenological predictions}\label{sect_results}

In the color allowed sector, based on an analysis of semileptonic decays and an
expansion in powers of $(\omega_0 -1)$, the
ratio in Eq.~(\ref{brpred1}) was previously predicted to be 
in the range $0.1-1.3$ in Ref.~\cite{Leibovich:1997em} and $0.35$ in 
Ref.~\cite{Neubert:1997hk}. In this paper, with the new power counting introduced at
maximum recoil, we have shown the 
ratio to be one at leading order. In fact we have obtained the same result even for the 
color suppressed channel. The main results of this paper at leading order are 
the equality of branching fractions and strong phases
\begin{eqnarray}\label{mainresult}
\frac {Br(\bar B \to D_2^{*}M)}{Br(\bar B \to D_1M)}=1\,, \qquad\quad 
\phi ^{D_2^*M} = \phi ^{D_1M} \,,
\end{eqnarray}
where $M=\pi,\rho ,K, K^*$ in the color allowed channel
and $M=\pi,\rho ,K, K_{||}^*$ in the color suppressed channel. This result in the color suppressed
channel is quite
unexpected from the point of view of naive factorization. In the color
suppressed channel the long distance operators in Fig.~\ref{fig_scet1}c,d give non-vanishing
contributions for kaons at leading order in $\alpha _s(Q)$ unlike the case of 
$M=\pi,\rho$. However, based on the same arguments~\cite{Mantry} given for the case of 
$B$-decays to ground state charmed mesons the long distance contributions to the color 
suppressed decays $\bar B^0 \to D_1^0\bar K^0$ and $\bar B^0 \to D_2^{0*}\bar K^0$ are equal
and the result still holds. For $K^*$'s the long distance contributions are equal only
when they are longitudinally polarized. 

Once data is available for the color suppressed channel we can construct 
isospin triangles analogous to
Fig.~(\ref{fig_isospin}). With $A_{0-}$ chosen as real, the strong phase
$\phi^{D**M}$
generated by the color suppressed channel $A_{00}$ through the soft functions $Q_{L,R}^{(0,8)}$ in 
Eq.~(\ref{hatsoft}), is identical for $D_1$ and $D_2^*$. The isospin angle 
$\delta $ which is related to $\phi $
through Eq.~(\ref{triangle}) is also the same for $D_1$ and $D_2^*$. Thus, at 
leading order we predict
the isospin triangles for $D_1$ and $D_2^*$ to identically overlap.

Recent data~\cite{Abe:2003zm,Aubert:2003hm} reports the ratio of branching fractions 
in the color allowed channel
\begin{eqnarray}\label{Belle}
\frac {Br(B^- \to D_2^{*0}\pi ^-)}{Br(B^- \to D_1^0\pi ^-)} = 0.79 \pm 0.11.
\end{eqnarray}
The deviation of this ratio from one, which will cause
the isospin triangles to no longer overlap, can be attributed to subleading
effects. The subleading effects shown to be suppressed by $\Lambda _{QCD}/Q$ 
are expected to give a 20\% correction, enough to bring agreement with current data.
Thus, our claim that subleading corrections are suppressed $\Lambda_{QCD}/Q$ is
in agreement with current data.


\section{Summary and Conclusions}\label{sect_concl}

In this paper we have presented a model independent analysis of two body 
$B$-decays to an excited charmed meson ($D_1$,$D_2^*$) and a light meson $M=\pi, \rho, K$.
The $b\to c$ flavor changing effective Hamiltonian was matched onto operators in
the soft-collinear effective theory (SCET) through a series of matching and running
steps. Factorization of the soft and collinear degrees of freedom was achieved in SCET
allowing us to use the tools of heavy quark symmetry (HQS) in the soft sector. The
combination of factorization with HQS lead to quantitative predictions relating the $D_1$
and $D_2^*$ amplitudes at leading order in $\Lambda _{QCD}/Q$ where $Q=\{m_b,m_c,E_M\}$.

The analysis closely paralleled the analysis for $B$-decays with ground state charmed
mesons $\bar B\to D^{(*)}M$~\cite{Mantry}. At leading order, the results of factorization,
generation of strong phases, and the $\Lambda _{QCD}/Q$ suppression of
color suppressed modes remain unchanged. Any differences show up only at the
non-perturbative scale at which the binding of hadrons occurs. Thus, the Wilson 
coefficients that arise in matching calculations are identical and only 
the non-perturbative functions experience a change. 
Another difference unique to color allowed decays to excited charmed mesons arises from the constraint
of HQS. The leading order amplitude is required to be proportional to $(\omega _0-1)$
which is numerically the same order as $\Lambda _{QCD}/m_c$. 
The study of semileptonic $B$-decays~\cite{Leibovich:1997em} near zero recoil 
suggests that the presence of powers of
$(\omega _0-1)$ could suppress the leading order amplitude. 
However, in this paper we have shown that maximum recoil is a special 
kinematic
point where the HQS constraints enter in a manner that preserves the SCET
power counting scheme with no suppression from $(\omega _0-1)$. This was done by
introducing a new power counting scheme for factors of $(\omega _0 -1)$ unique to maximum recoil kinematics.
We can safely rely on leading order predictions with corrections suppressed by at least
one power of $\Lambda _{QCD}/Q$. This was explicitly verified for contributions
from subleading semileptonic form factors.

At leading order we have shown the equality of the branching fractions $\bar B \to
D_1M$ and $\bar B \to D_2^*M$  and their strong phases.
Once data on the color suppressed channels becomes available we can construct
isospin triangles similar to those in Fig.~\ref{fig_isospin}.
Other phenomenological predictions similar to the ones in section VI of ~\cite{Mantry}
can be made can be made to leading order in $\alpha _s(\Lambda _{QCD}Q)$ by 
using tree level expressions for the jet functions $J$. 

Possible future work includes computing the remaining subleading corrections
to account for the 20\% deviation of the data from the leading order
predictions. The analysis could also be repeated for
$B$-decays to other excited charmed mesons such as $D_0^*$ and $D_1^*$.

\acknowledgments I would like to thank I.W.Stewart for suggesting this problem and 
for the many discussions and comments on the manuscript. I would also like to
thank Dan Pirjol for many useful comments and suggestions. This work is
supported in part by funds provided by the U.S. Department of Energy (D.O.E)
under the cooperative research agreement DF-FC02-94ER40818.




\end{document}